%% file: new_Manuscript.tex
\documentclass{bmcart}

\usepackage[utf8]{inputenc} 

\usepackage{CJKutf8} 
\usepackage{upgreek}  

\def\includegraphics{}

\usepackage{graphicx} 
\usepackage{makecell} 
\usepackage{amsmath}  
\usepackage{hyperref} 

\startlocaldefs
\endlocaldefs

\begin{document}

\begin{frontmatter}

\begin{fmbox}
\dochead{REGULAR ARTICLE} 


\title{Attention dynamics on the Chinese social media Sina Weibo during the COVID-19 pandemic}


\author[
   addressref={aff1},
]{\inits{HC}\fnm{Hao} \snm{Cui}}
\author[
   addressref={aff1},             
   corref={aff1},  
   email={kerteszj@ceu.edu}   
]{\inits{JK}\fnm{János} \snm{Kertész}}


\address[id=aff1]{
  \orgname{Department of Network and Data Science, Central European University}, 
  \street{Quellenstrasse 51},                     %
  \postcode{A-1100}                                
  \city{Vienna},                              
  \cny{Austria}                                    
}

{\sffamily {\small {\color{blue}{*}}Correspondence: {\color{blue}{kerteszj@ceu.edu}}}}

{\sffamily {\small Department of Network and Data Science, Central European University, Quellenstrasse 51, A-1100, Vienna, Austria}}


\end{fmbox}


\begin{abstractbox}

\begin{abstract} 

Understanding attention dynamics on social media during pandemics could help governments minimize the effects. We focus on how COVID-19 has influenced the attention dynamics on the biggest Chinese microblogging website Sina Weibo during the first four months of the pandemic.
We study the real-time Hot Search List (HSL), which provides the ranking of the most popular 50 
hashtags based on the amount of Sina Weibo 
searches. 
We show how the specific events, measures and developments during the epidemic affected the emergence of different kinds of hashtags and the ranking on the HSL. 
A significant increase of COVID-19 related hashtags started to occur on HSL around January 20, 2020, when the transmission of the disease between humans was announced. Then very
rapidly a situation was reached where COVID-related hashtags occupied 30-70\% of the HSL, however, with changing content. We give
an analysis of how the hashtag topics changed during the investigated time span
and conclude that there are three periods separated by February 12 and March
12. In period 1, we see strong topical correlations and clustering of hashtags; in
period 2, the correlations are weakened, without clustering pattern; in period 3,
we see a potential of clustering while not as strong as in period 1.
We further explore the dynamics of HSL by measuring the ranking dynamics and the lifetimes of hashtags on the list. This way we can obtain information about the decay of attention, which is important for decisions about the temporal placement of governmental measures to achieve permanent awareness. Furthermore, our observations indicate abnormally higher rank diversity in the top 15 ranks on HSL due to the COVID-19 related hashtags, revealing the possibility of algorithmic intervention from the platform provider.
\end{abstract}




\end{abstractbox}
%

\end{frontmatter}




\section*{1 Introduction}

In our times of information deluge the dynamics of public attention is of eminent importance from many aspects, including education, politics, marketing and governance. In the new media the flow of information has dramatically accelerated, leading often to rapidly changing public attention. At the same time these media provide unprecedented possibilities to study attention dynamics~\cite{Wu_Huberman_2007, Russel_Neuman_etal_2014} as they produce Big Data open for investigation. The microblogging service Twitter~\cite{Twitter} is particularly suited to provide the basis for quantitative studies on the dynamics of public attention as the content of the messages is available~\cite{Twitter_research}. Accordingly, Twitter data have been used to identify classes of dynamic collective attention~\cite{Lehmann_etal_2012}, investigate party-related activity and to predict election outcomes~\cite{Eom_et_al_2015}. Furthermore, Twitter data have served as the basis of modelling attention dynamics during pre-election time~\cite{Ko_etal_2014} or studying the relationship between public attention and social emotions~\cite{Pen_etal_2017}.

Public attention becomes a focal issue in times of crises like pandemics. As early as 2010, four years after it was launched, Twitter was shown to be an adequate, real-time content, sentiment, and public attention trend-tracking tool~\cite{Chew_Eysenbach_2010} and was used to study rapidly-evolving public sentiment with respect to the epidemic H1N1~\cite{Signorini_etal_2011}. The analysis of tweets enabled to quantify the difference between attention and fear and their distance-dependence in the case of the Ebola epidemic~\cite{van_Lent_etal_2017}. 
For the present pandemic COVID-19, Twitter studies on public attention have occurred, focusing on the perception of policies by the public~\cite{Zavarrone_etal_2020,Lopez_etal_2020}, fighting COVID-19 misinformation~\cite{pennycook_misinfo_2020}, and the psychological impacts associated with social media exposure during the pandemic~\cite{mental_covid, psy_covid}.

The first COVID-19 epicenter was in China where the service of Twitter is blocked, whereas its local substitute Sina Weibo, is very popular~\cite{Weibo}. Therefore, it is natural to use data from Weibo for similar purposes as was introduced earlier for Twitter in other countries. Scientists have already recognized that this microblogging service provides important insight into the function of the Chinese society~\cite{Tong_Zuo_2014,Nip_Fu_2015}.
Recently, some studies have appeared dealing with the public attention towards COVID-19 on Sina Weibo. The attention of Chinese netizens to COVID-19 was measured by analyzing keyword frequency in retrieved COVID-related posts from randomly sampled Weibo users~\cite{Yuner_posts} or hashtags on Sina Weibo Hot Search List (HSL)~\cite{yuxin_descr}, and was studied for identifying the sentiment and emotion trends of public attention~\cite{yuxin_descr, xiaoya_emotion}.



{ Social media search indices have shown correlations with the epidemic curve \cite{seo_corr, review_corr, Thomas_corr, lei_corr, cuilian_covidcorr} and have been used for prediction of the transmission of infectious diseases, such as cholera~\cite{cholera}, Ebola \cite{kui_ebola}, influenza~\cite{review_corr} and Middle East Respiratory Syndrome (MERS)~\cite{seo_corr}. For the recent COVID-19, online search trends such as Google Trends and Baidu Index have shown strong correlations with real-world cases and deaths~\cite{Thomas_corr, lei_corr, cuilian_covidcorr} and were used to predict the number of new suspected or confirmed cases~\cite{lei_corr, cuilian_covidcorr}. In these studies, the searched keywords on social media are mainly the symptoms and the names of the disease. Nevertheless, there are various aspects of the pandemic that influence the society, including the measures and regulations given by the government, the scientific knowledge provided by healthcare experts, aspects related to frontline doctors and nurses, the donation behaviors between countries, the remote working new norm and so on. In this study, we also pose the so far less studied question about the temporal correlation patterns between different COVID-related topic categories with the changing world situations of COVID-19.}

The Sina Weibo Hot Search List (HSL) depicts the popularity ranking of the top 50
hashtags in real-time, 
according to an algorithm in which the number of searches on the hashtags is dominant. The ranking dynamics on Weibo enables us to investigate the dynamics of popular attention in such details which is not available on other platforms. We will focus on the temporal evolution of this ranking in order to study the attention dynamics.

Recently, ranking dynamics has been studied widely from sports~\cite{Blumm_etal_2012,Criado_2013,Morales_etal_2016} to scientists, journals or companies~\cite{Blumm_etal_2012}.  There are stable rankings (word frequencies) with little or no changes in the ranks~\cite{stable_rank} and there are volatile ones with vivid dynamics (mentions of Twitter hashtags)~\cite{Blumm_etal_2012}. Clearly, Weibo HSL belongs to the latter with rich dynamic properties. This ranking provides a proxy for the attention preferences of Weibo users enabling the quantification of the dynamics thereof, which reflects the changes in the attention due to events and measures.  
Instead of capturing the public attention using segmented key words from retrieved posts, we adopted the ``rank diversity"~\cite{Morales_etal_2016} measure on Weibo HSL to describe the rank dynamics of real-time popular hashtags which can serve as an indication of public attention in the Sina Weibo system. To the best of our knowledge, we are the first in relating the ranking dynamics of social media hot topics to the public attention dynamics. 
To see the effect of the pandemic on the public attention we compare the situation
before the break-out of COVID-19 
with the period during the pandemic up to the time when consolidation sets in. 

Collective attention towards online news decays with time due to the fade of novelty and attractiveness in the competition with the other news~\cite{Wu_Huberman_2007}. Recent studies show a peak of collective attention towards COVID-19 in late January 2020 and a subsequent collapse, in terms of the dynamic behavior of words used on Twitter~\cite{alshaabi2020worlds, dewhurst2020divergent}.
To further understand the less studied attention decay mechanism and the decay pattern with time, we describe the public attention decay towards COVID-19 based on the dynamics of the ranks and the durations of COVID-related hashtags on the Weibo HSL.
  



In this study, we focus on the HSL of Sina Weibo during the period of COVID-19, which provides rich data about public attention and its dynamics in China. Based on that data, we have been able to track the evolution of public attention in different periods during the pandemic, follow how the attention of the population shifted from one group of topics to another and study the changing correlation patterns of different COVID-related topic categories with the evolving COVID-19 situation in the world. During our studies we have discovered signatures of the possible algorithmic control on this social media platform.
Understanding the dynamics of public attention on social media promotes instant and effective communications among governments, health experts and the public, helping the government to monitor public opinion, maintain the stability of the society as well as develop and deliver more effective measures to minimize the effects of the pandemic.


The paper is organized as follows: in Section 2, we provide background information on Sina Weibo real-time Hot Search List (HSL) and methodologies on quantifying attention dynamics. In Section 3, we present our results on attention dynamics, correlations between different types of hashtags, and attention decay. In Section 4, we discuss and summarize the results. 

\section*{2 Materials and methods}

\subsection*{\textbf{2.1 Sina Weibo real-time Hot Search List (HSL)}}

Sina Weibo is the biggest Chinese microblogging website, with MAU (monthly active users) reaching 550 millon and DAU (daily active users) 241 million in March 2020~\cite{Weibomau}. Instead of using one hashtag at the beginning like Twitter, topics on Weibo are confined in double hashtags one at the beginning and one at the end of the topic description, for example, \#Pneumonia of unknown cause detected in Wuhan\#. In this paper, we use hashtag as the synonym of topic, we refer a hashtag as the content contained within the double \#s. A hashtag becomes popular at a given time as it is used in many tweets, gains a large number of searches, likes and discussions by the users. The Hot Search List on Weibo is a section that displays the 50 most popular hashtags in real-time. The hashtags on the HSL, together with their ranks and search volume indices are updated every minute~\cite{Weibointro} and new popular hashtags may emerge and others vanish. The search volume index is a comprehensive measure which takes into account multiple dimensions such as the number of searches in Sina Weibo and the quality of the user accounts involved in the search, for the aim of preventing manipulated fake popularity~\cite{Weiboindex}. The third and sixth ranks on the HSL are sometimes 
occupied by promoted advertisements labeled with the character ``\begin{CJK*}{UTF8}{gbsn}荐\end{CJK*}"~\cite{Weiboads} (meaning recommendation).

\subsection*{\textbf{2.2 Data}}

We took data from Weibo HSL to study attention dynamics as it captures vibrant real-time change of public attention. Due to the random existence of one or two commercial advertisements at the third and the sixth ranks, in order to get a constant length of non-advertisement hashtags on the HSL at each timestamp, we removed all the hashtags labeled with ``\begin{CJK*}{UTF8}{gbsn}荐\end{CJK*}", re-ranked the original HSL and took the top 48 hashtags for each timestamp. All the HSL we mentioned later in this paper mean the re-ranked HSL with 48 ranks. We directly downloaded the data from the HSL with a frequency of every 5 minutes from December 16, 2019 to April 17, 2020. There are in total 26022 hashtags and 9120 of them are related to the aspects of COVID-19. To relate social media contents with real-life pandemic situation in Mainland China, we collected the daily number of infections, deaths, and recoveries from the official website of National Health Commission of China~\cite{nhs}. In the following subsection we explain how we identified the different categories of hashtags. 

\subsection*{\textbf{2.3 Topical classification and correlations}} 

\begin{figure}[htbp]
	\begin{center}
		\includegraphics[scale=0.5]{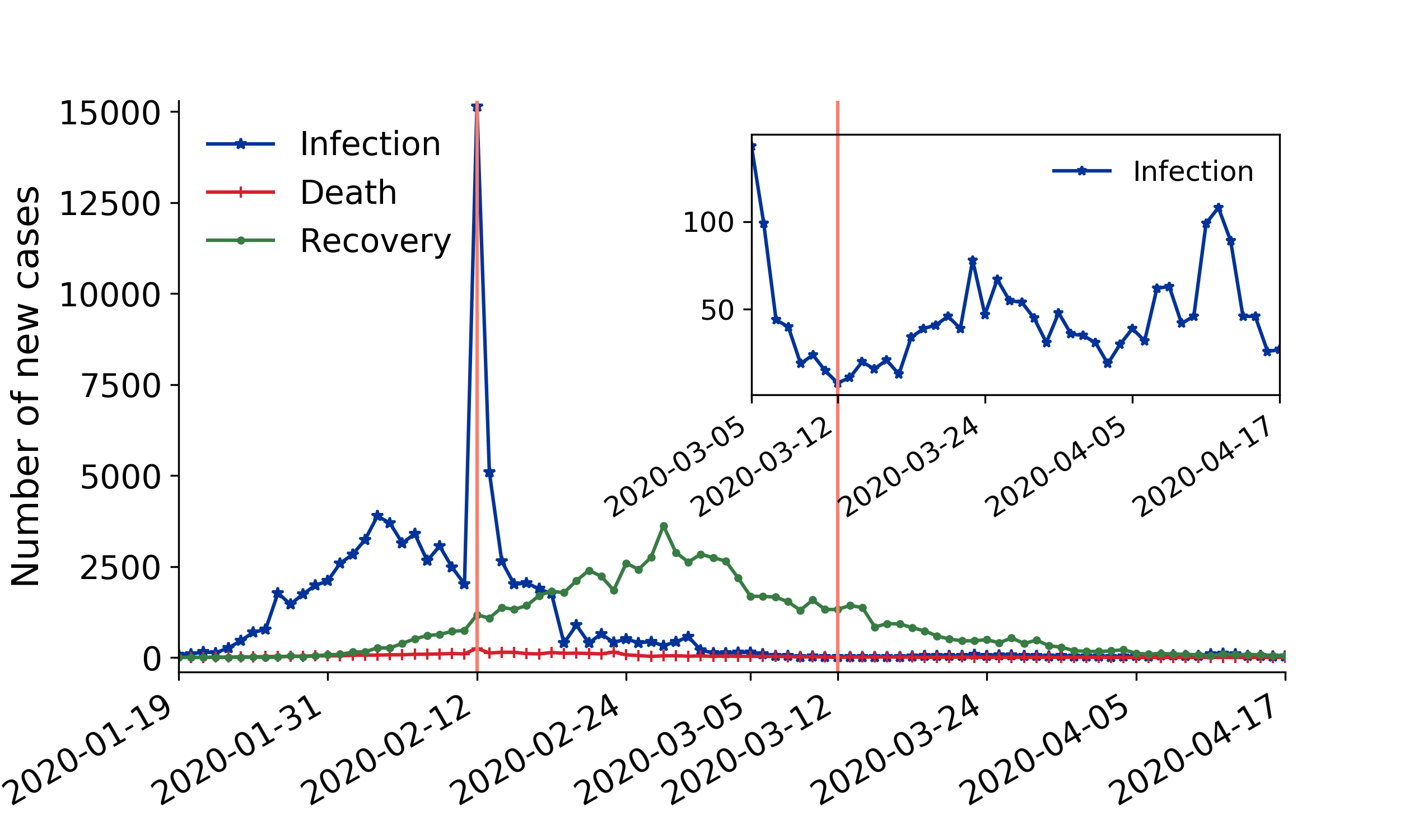}
	\end{center}
	\caption{COVID-19 daily infection, death and recovery in Mainland China. The inset enlarges the tail of the infection curve. Three periods after the outbreak on January 19 are separated by the highlighted peak and local minimum.}
	\label{fig:fig1}
\end{figure}

Fig. \ref{fig:fig1} shows the number of daily infections, deaths and recoveries in Mainland China. The number of daily infections and deaths have a sharp peak on February 12 due to the adoption of new diagnostic criteria~\cite{Feb12}. The decreasing trend of daily infections since the peak turned to increasing after March 13, as a result of the rising number of imported coronavirus cases from abroad~\cite{importcase}. We will argue that there are three periods to be distinguished after the outburst of COVID-19 around January 19, separated by the maximum and local minimum of the daily number of infections on February 12 and March 12, respectively. 


To study the public attention towards COVID-related information, we first extracted hashtags which encompass all aspects of COVID-19 and classified them based on geographic regions and the exposure order under the pandemic into three categories: Mainland China, East Asia outside of Mainland China and Other Countries outside of East Asia. With a focus on COVID-hashtags related to Mainland China, we manually classified them based on semantic meanings into the following seven disjoint sub-categories. The Bad News category comprises hashtags on confirmed infections and deaths in different regions of Mainland China as well as shortages of essential supplies. The Good News category consists of news on cases of recovery, sufficiency of supplies, and decrease in daily infections or deaths. The Regulations category consists of authority announcements of national, regional, institutional laws, rules and regulations associated with public behaviors and concerns during the pandemic. The Life Influence category contains hashtags that reflect the pandemic influence on the aspects of citizen lives. The Front Lines category includes hashtags related to the lives of front line workers (mainly doctors and nurses) and their interactions with patients in hospitals. The Science category incorporates scientific understandings of the virus properties, vaccine development, and ways for public protection given by authoritative doctors. The Supports category takes into account hashtags on worldwide donations and emotional supports. For ambiguous cases which contain information of more than one category, our classifications were based on the focus of the main subject. Due to the syntactic-semantic complexity of Chinese language, the classifications were made by two independent annotators. Final consensus was reached in case of disagreement. The Mainland China sub-categories are summarized in Table 1 together with examples. The full list of COVID-related hashtags is available in the dataset, which we have made public (see declaration at the end of the paper).

To further understand how the Mainland China related COVID-hashtags are correlated with each other and with the daily number of infections/deaths/recoveries in the three separated time periods, we measured the Pearson's correlations between the seven series of daily number of new hashtags in each of the sub-categories defined above, together with the three series of daily number of infections/deaths/recoveries. The correlation of these ten time series are calculated using the percentage change between the current and the prior element instead of the actual value in order to reduce the effect of the trend which can cause spurious correlations. For time series category $X = \{X_{t_i}: t_i \in T, i = 1,2, ...n\}$ and category $Y = \{Y_{t_i}: t_i \in T, i = 1,2, ...n\}$, where $T$ is the time index set, the Pearson's correlation is calculated using the percentage change series $\tilde{X} = \{ \frac{X_{t_{i+1}} - X_{t_i}}{X_{t_i}}, t_i \in T, i = 1,2, ...n \}$ and $\tilde{Y} = \{ \frac{Y_{t_{i+1}} - Y_{t_i}}{Y_{t_i}}, t_i \in T, i = 1,2, ...n \}$.

\begin{table}[!ht] 
\caption{Mainland China COVID-hashtag details. A summary of the example hashtags in each sub-category of Mainland China category and the number of hashtags in different time periods.}
\scalebox{0.783}{
      \begin{tabular}{ccccc}
        \hline
        Category   & Examples & Period 1  & Period 2 & Period 3 \\ \hline
        Bad News &  \makecell{
        \begin{CJK*}{UTF8}{gbsn} \#全国累计确诊新冠肺炎66492例\#\end{CJK*}  \\
                                 (\#National cumulative confirmed COVID-19\\ cases reach 66492\#) \\   
        \begin{CJK*}{UTF8}{gbsn} \#黑龙江聚集性疫情共48起发病194人\#\end{CJK*}\\
                                 (\#Heilongjiang in total 48 clustered epidemic\\ 194 infected cases\#)\\
        \begin{CJK*}{UTF8}{gbsn} \#武汉多家医院物资紧张\#\end{CJK*} \\
                                (\#Wuhan many hospitals supplies in shortage\#)}
 & 451 & 193 & 272 \\\hline 
        Good News & \makecell{
        \begin{CJK*}{UTF8}{gbsn}\#火神山医院累计治愈患者破千\#\end{CJK*}\\
        (\#Huoshenshan Hospital has cured over a thousand patients\#)\\
        \begin{CJK*}{UTF8}{gbsn}\#7省区现有确诊病例清零\#\end{CJK*}\\
        (\#7 provinces current infected cases down to zero\#)\\        
        \begin{CJK*}{UTF8}{gbsn}\#疫情形势出现3个积极变化\#\end{CJK*}\\
        (\#Epidemic situation shows 3 positive changes\#)}      
                                 &145  & 257  & 121\\\hline

        Regulations & \makecell{
        \begin{CJK*}{UTF8}{gbsn}\#上海地铁不戴口罩不得进站\#\end{CJK*}\\
        (\#Shanghai metro station not allowed to\\ enter without wearing a mask\#)\\
        \begin{CJK*}{UTF8}{gbsn}\#疫情影响严重的地区可增发生活补助\#\end{CJK*}\\
        (\#Additional living allowances can be issued in areas \\ severely affected by the epidemic\#)\\        
        \begin{CJK*}{UTF8}{gbsn}\#非疫情严重国家进京者居家观察14天\#\end{CJK*}\\
        (\#Home observation for 14 days for visitors to enter Beijing \\ from non-severe epidemic countries\#)}               
        & 318 & 325 & 633\\\hline

        Life Influence & \makecell{
        \begin{CJK*}{UTF8}{gbsn}\#武汉市民江滩唱起国歌\#\end{CJK*}\\
        (\#Wuhan citizens sing national anthem at the River Beach\#)\\ 
        \begin{CJK*}{UTF8}{gbsn}\#疫情期间点外卖指南\#\end{CJK*}\\
        (\#Guide to ordering takeout during the epidemic\#)\\     
        \begin{CJK*}{UTF8}{gbsn}\#一季度民航业亏损398亿\#\end{CJK*}\\
        (\#Civil aviation industry suffered a loss\\ of 39.8 billion in the first quarter\#)}                                
        & 310 & 371 & 649 \\\hline

        Front Lines & \makecell{
        \begin{CJK*}{UTF8}{gbsn}\#钟南山等专家连线武汉ICU团队\#\end{CJK*}\\
        (\#Zhong Nanshan and other experts connected\\ to the Wuhan ICU team\#)\\                                
        \begin{CJK*}{UTF8}{gbsn}\#护士握手呼唤79岁新冠患者\#\end{CJK*}\\
        (\#Nurse shakes hands and calls 79-year-old COVID-19 patient\#)\\ 
        \begin{CJK*}{UTF8}{gbsn}\#方舱医院收治第一批患者现场\#\end{CJK*}\\
        (\#Site of Fangcang shelter hospital taking the first batch of patients\#)}                                        
        & 251 & 329 & 347\\\hline

        Science & \makecell{
    \begin{CJK*}{UTF8}{gbsn}\#各年龄段人群普遍易感新冠病毒\#\end{CJK*}\\
        (\#People of all ages are generally susceptible to coronavirus\#)\\
        \begin{CJK*}{UTF8}{gbsn}\#口罩的正确使用方法\#\end{CJK*}\\
        (\#The correct use of masks\#)\\
        \begin{CJK*}{UTF8}{gbsn}\#如何区分感冒流感和新冠肺炎\#\end{CJK*}\\
        (\#How to distinguish between flu and COVID-19\#)}
        & 180  &  170  & 123\\ \hline

        Supports & \makecell{
    \begin{CJK*}{UTF8}{gbsn}\#汶川村民自发支援武汉100吨蔬菜\#\end{CJK*}\\
        (\#Wenchuan villagers spontaneously \\support Wuhan 100 tons of vegetables\#)\\                                
    \begin{CJK*}{UTF8}{gbsn}\#欧盟对华运送12吨急需物资\#\end{CJK*}\\
        (\#EU sends 12 tons of urgently needed supplies to China\#)  \\
    \begin{CJK*}{UTF8}{gbsn}\#武汉给援汉医疗队全员的感谢信\#\end{CJK*}\\
        (\#Thank you letter from Wuhan to all \\ members of the medical aid team\#)}    
        & 151 & 144  & 116\\

        \hline
      \end{tabular} }
\end{table}

\subsection*{\textbf{2.4 Attention dynamics}}

One natural measure of social media attention towards a topic category is the quantity of the related hashtags. The growing pattern of the cumulative number of hashtags on the HSL with time reflects the dynamics of the public attention. We separately measured the growth of the cumulative number of all hashtags and all COVID-related hashtags that ever appeared on the HSL in our observation period. To understand how much COVID-information occupies the HSL at each timestamp, we constructed the historical ratio trajectory of the COVID-related hashtags on the HSL since the first COVID-hashtag \begin{CJK*}{UTF8}{gbsn} \#武汉发现不明原因肺炎\# \end{CJK*} (\#Pneumonia of unknown cause detected in Wuhan\#) appeared on December 31, 2019. 

\subsubsection*{2.4.1 Lifetime duration}

The lifetime duration of a hashtag on the HSL indicates the ability of obtaining persistent attention from the public. We quantified the duration (continuous existence on the HSL) of a hashtag with $\uptau$:
$$\uptau = \tau_1-\tau_0,$$
where $\tau_0$ is the timestamp of the first 
and
$\tau_1$ is the timestamp of the last appearance of a hashtag on the 
HSL.

We compared the duration of the hashtags across various categories and different time scopes. We compared the duration of the hashtags before the outbreak on January 19, all COVID-related hashtags, and non-COVID hashtags after the outbreak. To ensure complete life cycles of the hashtags, we took all hashtags whose first arrivals on the HSL are between December 19, 2019 and January 18, 2020 as the sample for hashtags before the pandemic, which includes 6161 in total. Similarly, we took all COVID-hashtags whose first arrivals are no later than April 14, with a total number of 8808. For the non-COVID hashtags after the outbreak, we took a random sample of all non-COVID hashtags with the same size as the COVID sample. Hashtags that reappeared after disappearing from the HSL were excluded from our calculation. To understand the overall attention variation towards COVID-hashtags with time, we investigated the daily value of their cumulative average duration. We denote $D_{j}$ as the cumulative average of duration from December 31, 2019 (day 0) until day $j$. $D_{j}$ is calculated as follows:
\begin{equation}
    D_j = \frac{1}{|S(j)|}\sum_{i=0}^j \sum_{\alpha \in S(j)} d_{i}^{\alpha}
\end{equation}
where $d_{i}^{\alpha}$ is the duration of hashtag $\alpha$ whose first appearance was on day $i$. $S(j)$ is the set of all the hashtags whose first appearance is in the interval $[0, j]$.


\subsubsection*{2.4.2 Ranking}

The changes in the ranking patterns of the hashtags at different time periods reflect the general public attention dynamics. Rank diversity $d(k)$~\cite{Morales_etal_2016} is defined as the number of distinct elements in a complex system that occupy the rank $k$ at some point during a given length of time.   
Rank diversity is known to give characteristic profiles for different types of systems; e.g., in open systems (where only the top part of the competing items is ranked) behaves differently from closed systems (where all the items are ranked)~\cite{Morales_etal_2016, stable_rank}.
In this paper, we use rank diversity to measure the number of different hashtags occupying a given rank on the HSL over a given length of time, and thus obtain overall information on the total dynamical trend of the hashtags on the HSL. We normalize the rank diversity value by the total number of unique hashtags that have appeared on the HSL in a given time interval.
We compared the rank diversity in the 48 ranks on the 
HSL before the outbreak and during the different periods after the outbreak, with and without COVID-19 hashtags. 

The public attention towards a hashtag can also be indicated by its highest rank during the lifetime on the HSL. The highest rank of a hashtag reveals its highest ability and achievement when competing for attention with the other hashtags. We studied the highest rank distribution of the classified COVID-hashtags and compared the results with the hashtags before the outbreak as well as the non-COVID hashtags after the outbreak (SI). To understand the overall highest rank variation towards COVID-hashtags with time, we investigated the daily value trajectory of their cumulative average highest rank. We denote $H_{j}$ as the cumulative average of highest rank from December 31 (day 0), 2019 until day j. $H_{j}$ is calculated as follows:

\begin{equation}
H_j = \frac{1}{|S(j)|}\sum_{i=0}^j \sum_{\alpha \in S(j)} h_{i}^{\alpha}
\end{equation}
where $h_{i}^{\alpha}$ is the highest rank of hashtag $\alpha$ whose first appearance was 
on day $i$. $S(j)$ is the set of all the hashtags whose first appearance was in the interval $[0, j]$.

\section*{3 Results}

\subsection*{\textbf{3.1 Statistics and categorization of hashtags}} 

\begin{figure}[htbp]
	\begin{center}
		\includegraphics[scale=0.37]{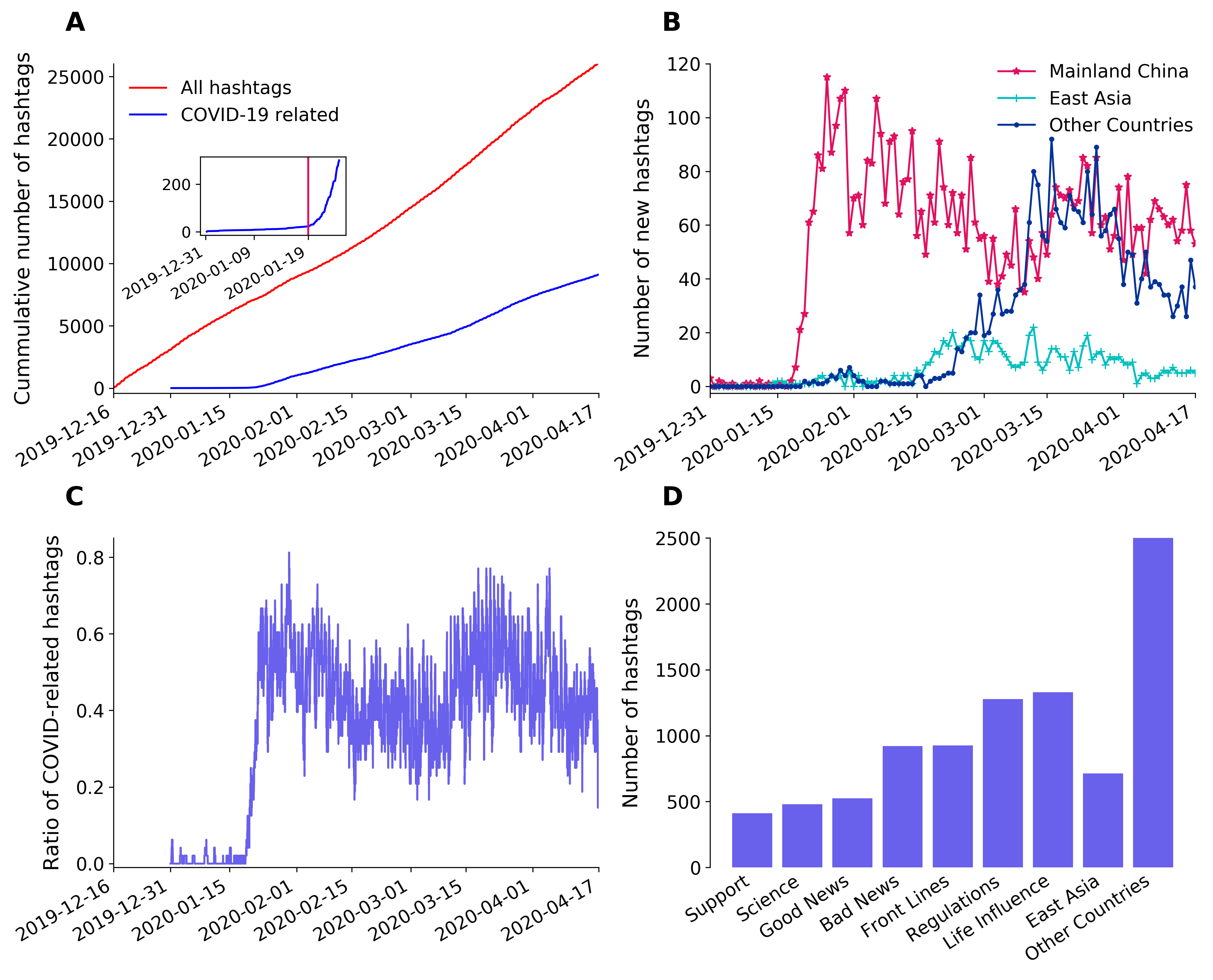}
	\end{center}
	\caption{Overview of COVID-hashtags on Weibo 
		re-ranked Hot Search List (HSL) throughout the pandemic. (A) Cumulative number of all hashtags and all COVID-hashtags with time. The inset indicates rapid increase in COVID-related hashtags starting from January 19 marked by a vertical red line. (B) Daily new COVID-hashtags on Mainland China, East Asia outside of Mainland China and Other Countries outside of East Asia. (C) Ratio of COVID-hashtags on the 
		HSL at each timestamp. (D) Distribution of all COVID-hashtags by categories.}
	\label{fig:fig2}
\end{figure}

The cumulative number of new hashtags on HSL grows approximately linearly (see Fig. \ref{fig:fig2} (A)), indicating a nearly constant attention capacity and need for news of the users. Closer inspection tells, however, that the rate of new hashtags decreases between January 10 and February 12 followed by an increased rate until 
March 28
after which the original slope of $225\pm 4$ new hashtags/day sets in. We attribute this change in the slope to the effect of COVID-related hashtags.

The first COVID-related hashtag appeared on the HSL on December 31, 2019, followed by only a few ones in the following week. As the first death case occurred on January 11, second one occurred on HSL on January 16 and more infected cases detected in other cities in China as well as in the surrounding Asian countries, rumours and scared emotions about the unknown pneumonia were permeating in the society and the number of daily COVID-related hashtags started to increase rapidly on January 19. On January 20, Chinese authorities announced to the public that the new coronavirus is transmissible between humans.    

From our point of view the period until January 19 can be considered as pre-COVID. During that time at most three COVID-related hashtags per day have occurred on the HSL and the cumulative number of different hashtags on HSL has grown approximately linearly with an unaltered slope (see Fig.~\ref{fig:fig2} (A)). 
Around January 19 
the number of COVID-related hashtags started growing and, at the same time, the overall growth of the total number of hashtags slightly decreased, indicating that the new COVID-related hashtags stay longer on HSL as compared with those before the outbreak. This results in a decrease of the total number of new hashtags per unit time on HSL. After January 19, a rapid increase can be observed in the number of COVID-hashtags (see the inset of Fig.~\ref{fig:fig2} (A)). This has, finally, also an effect on the total cumulative number of hashtags resulting in an increased slope in Fig.~\ref{fig:fig2} (A).

Fig. \ref{fig:fig2} (B) shows the cumulative number of geographically categorized COVID-hashtags with Mainland China, East Asia outside of Mainland China, and Other Countries outside of East Asia as categories. The Mainland China category starts to rise rapidly from January 19, reaches a peak in the following week, and then gradually drops with a few rebounds. The second peak and the decline of Mainland China category is intertwined with the trajectory of the Other Countries category in mid-March. The East Asia category remains at a relatively low level throughout the pandemic. 


COVID-19 was first observed in east Asia, with Mainland China being the hardest-stricken region, followed by places with growing infections such as South Korea, Diamond Princess cruise ship and Japan. The epicenter of COVID-19 later shifted to Europe and the rest of the world as the situation mitigated in east Asia. The results depicted in Fig. \ref{fig:fig2} (B) follow these events closely, confirming the role of the real-time HSL on Weibo as a reflection of the real world. Unsurprisingly, the upward and downward trend periods of Mainland China and Other Countries coincide with Fig. \ref{fig:fig2} (C), where the ratio of COVID-related hashtags on the HSL at each timestamp is displayed. The swift third peak on April 4 in Fig. \ref{fig:fig2} (C) is due to the national Qingming Festival (also known as the Tomb-Sweeping Day), where the victims who died in the COVID-19 pandemic were mourned. The dynamics of the COVID-related hashtags on the HSL demonstrates vibrant generations of newly created COVID-19 hashtags about the relevant up-to-date events around the world. Fig. \ref{fig:fig2} (D) shows the distribution of the hashtags in the sub-categories of Mainland China category along with East Asia and Other Countries. Among the seven sub-categories that belong to Mainland China, Support, Science, and Good News have relatively fewer hashtags, compared with Front lines, Life Influence, Bad News, and Regulations.

\subsection*{\textbf{3.2 Periodization and correlations}}

\begin{figure}[htbp]
	\begin{center}
		\includegraphics[scale=0.43]{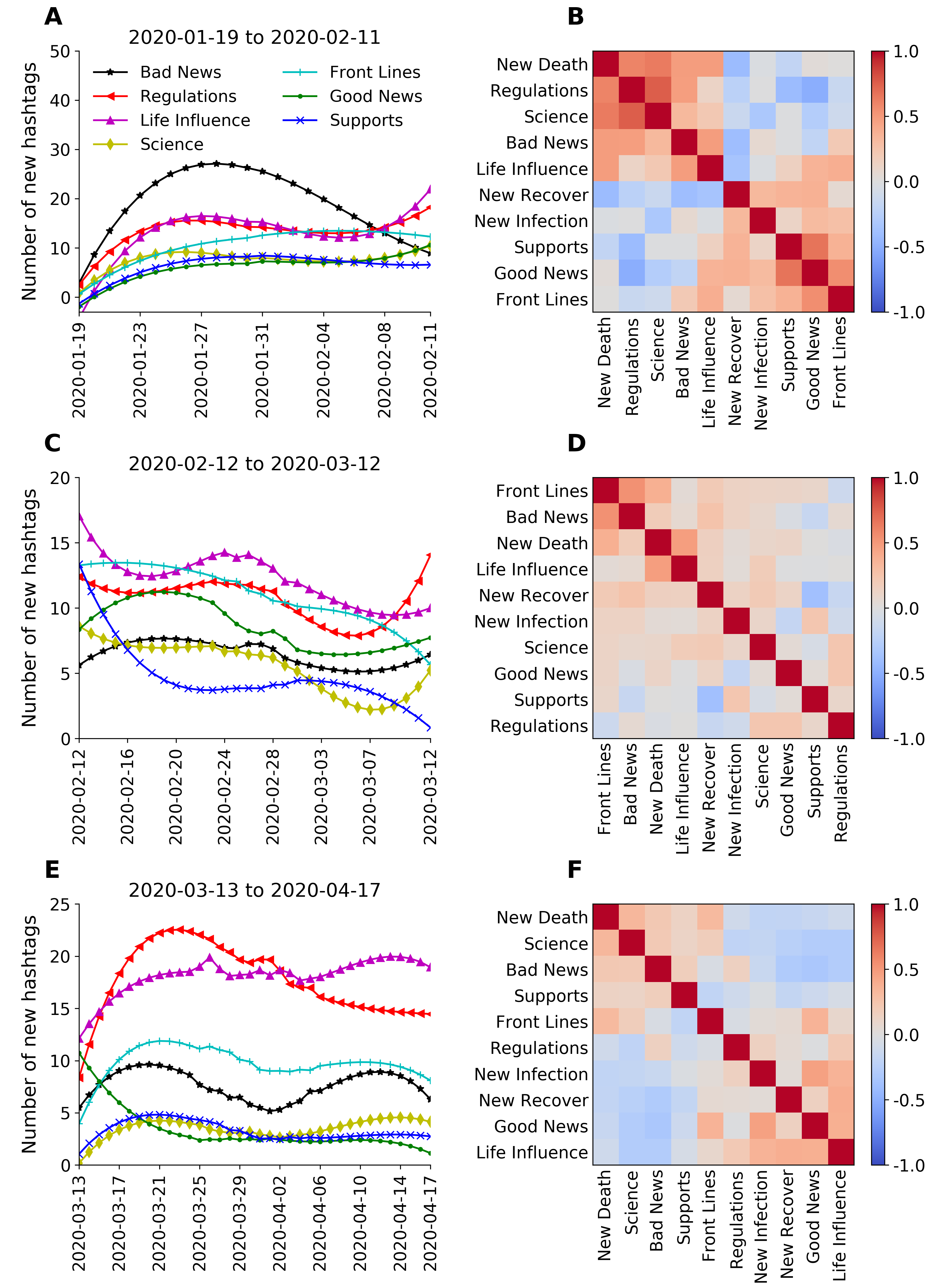}
	\end{center}
	\caption{Time series of daily new hashtags (smoothened by a Savitzky-Golay filter~\cite{smoothen} with polynomial order 3) from the sub-categories of Mainland China COVID-hashtags and their correlation matrices with daily new infections, deaths, and recoveries, in the three periods after the outbreak.}
	\label{fig:fig3}
\end{figure}

Fig. \ref{fig:fig3} illustrates the attention dynamics of the sub-categories of Mainland China by showing the quantity variations in Fig. \ref{fig:fig3} (A) (C) (E), paired with their correlation matrices with daily infections, deaths, and recoveries in Fig. \ref{fig:fig3} (B) (D) (F). 
As noted above, we have identified three periods in the investigated time interval: The first period is January 19 -- February 11, separated by the huge peak in Fig. \ref{fig:fig1} from the the second one (February 12 -- March 12). The third period (March 13 -- April 17) 
is separated from period 2 by the second vertical line where the number of new infections has a local minimum (Fig. \ref{fig:fig1} inset).

In Table 1, we show the number of hashtags related to Mainland China in the different categories for the three periods. In Fig. \ref{fig:fig3}, we show that the daily emergence of the categorized COVID-hashtags is dominated in the first two periods by Bad News, with increasing and decreasing trends in period 1 and period 2, respectively. In period 3, the categories Regulations, Life Influence, and Front Lines receive more attention as compared to the rest of the categories. Here the consistently high values in Regulations and Life Influence could result from the worsening world pandemic situation along with the rise of the imported infected cases in Mainland China, necessitating the establishment of measures to handle it. The categories of the Mainland China COVID-hashtags move with the number of infections and deaths in the world. 


The patterns of the Pearson's correlation matrix of the ten time series reflect temporal structure with the three periods. 
Fig. \ref{fig:fig3} (B) shows a positive correlation block structure. There are strong correlations between New Death, Regulations, Science, and Bad News (upper left block) as well as between Supports, Good news and Front Lines (lower right block) and there is considerable anti-correlation between the two blocks.  
Fig. \ref{fig:fig3} (D) (period 2) exhibits much weaker correlations, in fact, very few elements of the matrix reach values beyond the noise level (see SI). 
Exceptions are new strong correlations between New Death and Front Lines, as well as Bad News and Front Lines. 
In the third period (Fig. \ref{fig:fig3} (F)) the block structure gets again more pronounced, though not as pronounced as in the first period. Note that the categories had to be rearranged in order to achieve this structure.
The major change is that Supports/Front Lines and Life influence/Regulations have exchanged positions. In period 1, the Bad News (mainly infections and deaths) of domestic cases in Mainland China were flooding, this lead to the urgent establishment of regulations, which caused life influences. In period 3, the domestic situation was under control, therefore, the Bad News in Mainland China were mainly caused by the worsening international situation  (infections/deaths and Chinese coming back from abroad). Then the Regulations and corresponding Life Influences towards these issues were not anymore strongly associated with domestic deaths.

\subsection*{\textbf{3.3 Rank diversity and hashtag dynamics}} 

\begin{figure}[htbp]
	\begin{center}
		\includegraphics[scale=0.4]{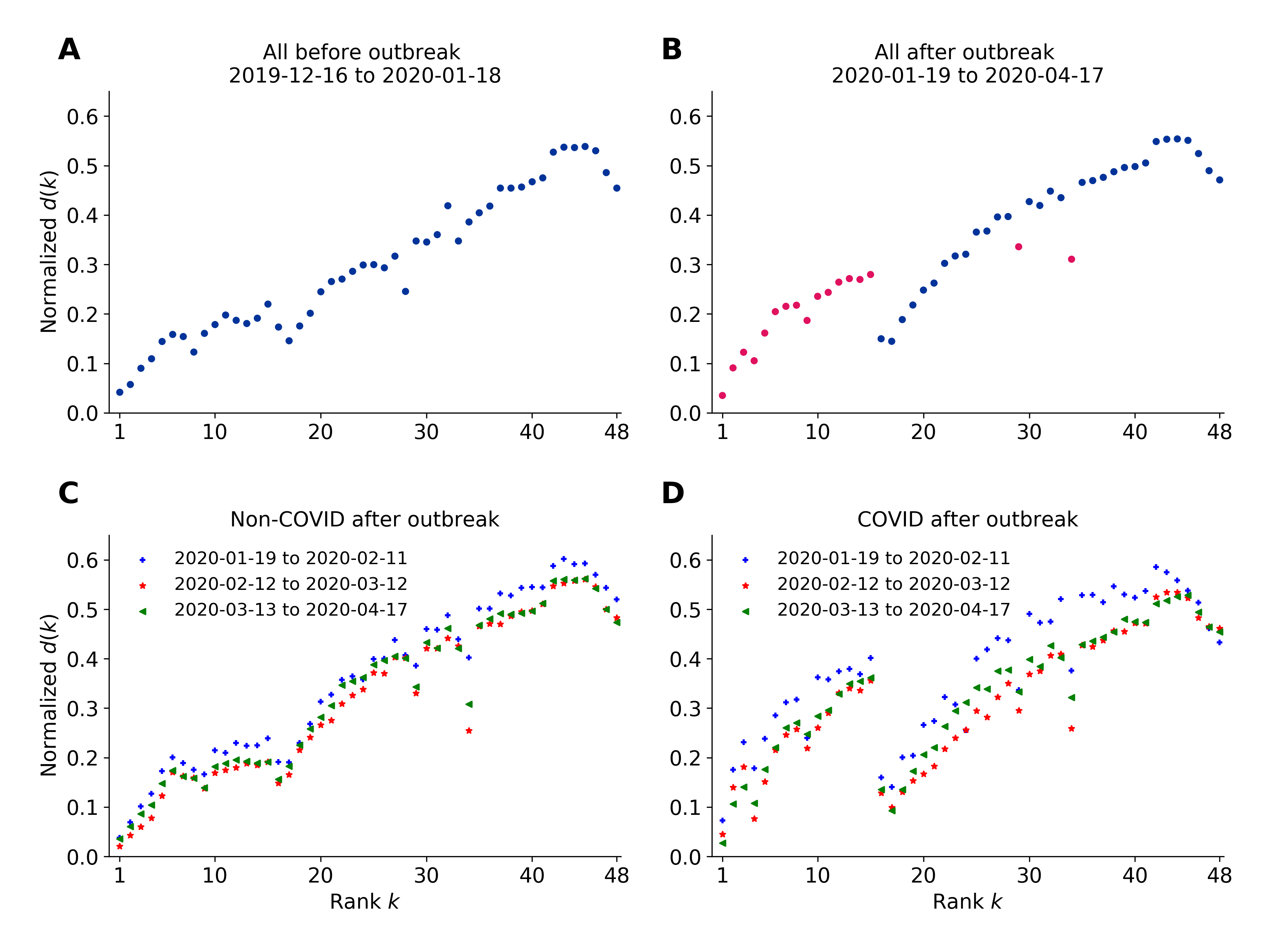}
	\end{center}
	\caption{Rank diversity $d(k)$ of the 48 ranks on the HSL before and after COVID-19 outbreak. (A) Rank diversity taking all hashtags in our observation period before the outbreak, approximately linear except the head and tail parts, with small fluctuations. (B) Rank diversity taking all hashtags after the outbreak, with strange points colored in red. A large gap occurs after the top 15th rank. (C) Rank diversity taking all non-COVID hashtags in the three periods after the outbreak, strong resemblances with (A). (D) Rank diversity taking all COVID-hashtags in the three periods after the outbreak. The result in period 1 is higher than period 2 and period 3, revealing a more dynamic change of the hashtags appeared on the HSL. The gap after rank 15 is more severe compared to (B).}
	\label{fig:fig4}
\end{figure}

What is the effect of COVID-19 on the ranking dynamics? Fig. \ref{fig:fig4} shows a comparison of the rank diversity in the top 48 ranks taking non-COVID and COVID hashtags in different periods. Striking differences are observed between the rank diversity plots before and after the outbreak. As Fig.~\ref{fig:fig4} (A) suggests, the rank diversity plot before the outbreak was approximately linear with moderate fluctuations. A clear gap emerges in the rank diversity after rank 15 in Fig.~\ref{fig:fig4} (B) during the COVID period. We recognize resemblances in the rank diversity plots before the outbreak and after the outbreak considering only non-COVID hashtags, except for the strange drops at ranks 29 and 34 in Fig. \ref{fig:fig4} (C). 
Comparing Fig. \ref{fig:fig4} (D) with Fig. \ref{fig:fig4} (B), the gap after rank 15 is larger in the rank diversity plot considering only COVID-related hashtags. The rank diversity plots for hashtags in period 1 surpass period 2 and period 3 with both non-COVID and COVID hashtags as depicted in Fig. \ref{fig:fig4} (C) and (D), while the difference is much higher in the latter case. 


Fig. \ref{fig:fig4} gives evidence that the COVID-hashtags cause the gap in the rank diversity plot after the outbreak. Taking the normalized rank diversity plot before the outbreak as a reference, a higher normalized rank diversity at a certain rank position represents a higher number of unique occurrences within the observation period, so that the COVID-related hashtags in the top 15 ranks change faster (with higher frequency) than normal. One possible explanation is that the COVID-hashtags kept emerging with higher frequency than before the outbreak and people payed much attention to these new hashtags. Additionally, when the flooding hashtags contained similar information such as the new infections and deaths in different cities or provinces of China, the public interest towards individual hashtags could drop quickly, resulting in a higher number of unique hashtags at certain ranks in unit time on HSL. This effect of higher rank diversity for higher ranks seems to be amplified by the algorithm leading to the observed gap.

Strange drops of rank diversity at ranks 29 and 34 can also be seen on our plots in Fig. \ref{fig:fig4}. As provided in SI, there are hashtags that stay at the ranks 29 and 34 for an unusually long time and then disappear from the HSL, indicating algorithmic intervention from Weibo. As one of the most popular and influential social media in China, Weibo might shoulder the responsibility during the global public health emergency to keep people informed about related news in China and around the globe, by means of changing the algorithm towards COVID-hashtags to promote crucial news and keep them updating in the top 15 positions and leave the list at rank 29 or 34. Our methods are sensitive enough to demonstrate this type of interventions. Therefore our observations reflect a combination of both spontaneous attention dynamics from the public and the controlled effects from Sina Weibo.

Rank diversity captures attention dynamics from the point of view of the overall dynamical rank movements of the hashtags on the HSL. It is interesting to follow the dynamics also from the aspect of the individual hashtags. The average highest rank of a category of hashtags on a given day is characteristic to the attention paid to that category. (Note, of course, that getting to the HSL expresses already considerable attention.) Similarly, the average duration is another measure of attention. However, in the latter case it should be mentioned that short duration can be caused by decaying attention to the general topic (in this case the hashtag is likely to be replaced by another from a different topic) or because of the heavy stream of new hashtags of the same topic. 

\begin{figure}[htbp]
	\begin{center}
		\includegraphics[scale=0.35]{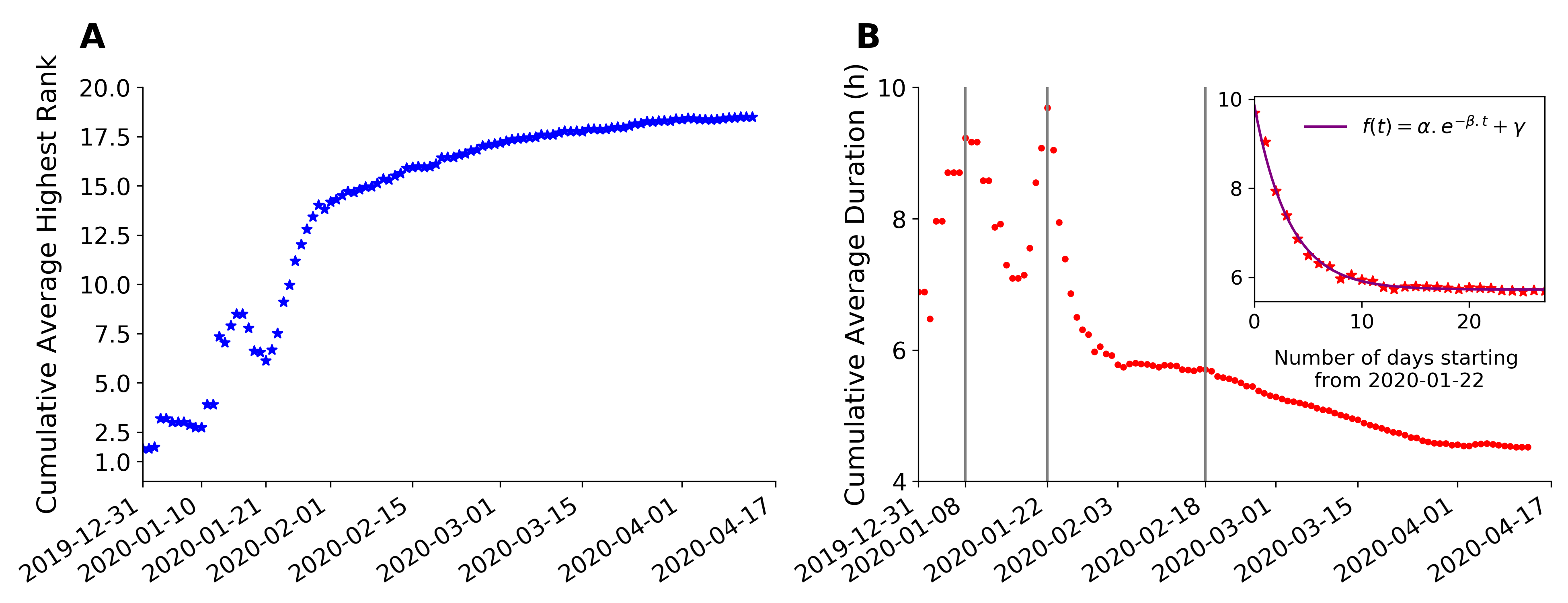}
	\end{center}
	\caption{Attention decay. (A) Cumulative average highest rank of COVID-hashtags whose first appearance was in the time interval since December 31, 2019. (B) Cumulative average duration (hours). The inset shows a three-parameter exponential fit ($\alpha = 4.13{\rm h}, \beta = 0.31 {\rm h/day}, \gamma = 5.72{\rm h}$) for the cumulative average duration decay after January 22, 2020.}
	\label{fig:fig5}
\end{figure}

How do the average highest rank and average duration accumulate with time? As Fig. \ref{fig:fig5} (A) shows, the cumulative average highest rank, $H_j$ is initially at a top rank, indicating that the first few hashtags about the unknown pneumonia received a huge amount of attention from the public. As more COVID-related hashtags occurred, $H_j$ becomes lower, with a rapid change at the beginning and a slower change later, separated by around January 30. This is due to the rapidly increasing number of COVID-related hashtags and the limited number of ranks on HSL.
In Fig. \ref{fig:fig5} (B), the first peak of the cumulative average duration, $D_j$ is on January 8, when the hashtag was posted that eight patients infected by the unknown pneumonia recovered from hospital. Then the $D_j$ decreases first and then increases again, reaching the second peak on January 22, after which the increasing daily new hashtags with short durations started to play a greater role than the few hashtags with long durations.

The fast decay of $D_j$ in the period between January 22 and February 18 (see the inset in Fig. \ref{fig:fig5} (B)) was fitted by an exponential function:
\begin{equation}  \label{eq:2}
    f(t) = \alpha e^{-\beta.t}+\gamma,
\end{equation}
with $\alpha = 4.13{\rm h}, \beta = 0.31 {\rm h/day}, \gamma = 5.72{\rm h}$. On February 18, hashtags of positive changes in the COVID situation started to appear on HSL. After that, the $D_j$ exhibits a slower and longer decay.

\section*{4 Discussion and summary} 

In this work, we have studied the public attention dynamics on the real-time Hot Search List (HSL) of the biggest Chinese microblogging website Sina Weibo under the influence of the COVID-19 pandemic. On the one hand, such study contributes to the understanding of the dynamics of public attention on social media and how it reflects the dynamics of the public thoughts and behaviors. On the other hand, identifying the online attention dynamics patterns and their relationship to events and measures during pandemic may contribute to its efficient management.

In order to follow the dynamics of public attention we have introduced sub-categories of COVID-19-related hashtags. Our results show diversification of the public attention after the outbreak on January 19, 2020 as indicated by changing frequencies of such hashtags in the different sub-categories. Moreover, the pattern of correlations with the real-world events and measures vary in three identified periods during the investigated time span. We conclude that at the beginning the dominant driving force of the public attention was the infection and death situation in Mainland China, with mainly domestic cases, while the international situation and the imported cases influenced the attention later. Our observations point towards the complexity of the attention patterns indicating that several components should be taken into account if such data are used to the prediction of the epidemic curve~\cite{lei_corr, cuilian_covidcorr}. 

Furthermore, we have shown that the cumulative average duration follows exponential decay immediately after the attention peak in the pandemic, but a slower decay for a longer time. 
The exponential decay suggests that the speed of governmental response is crucial in the early pandemic phase. This exponential decay sets a scale for the governments within which it should take quick actions and publish crucial measures and regulations to control the pandemic, healthcare experts should deliver scientific knowledge to inform the public how to protect themselves efficiently. 

The attention towards COVID-hashtags decayed as the circumstances in China got better. Nevertheless, the attention was influenced by the world pandemic situation which kept changing, hence the decay of public attention on the Chinese social media Weibo has become less clear cut. In any case, targeted and timely stimulus should be given to keep the attention and awareness of the public throughout the pandemic to prevent future waves of COVID-19.

In this paper we have made the first step to relate the ranking dynamics of hot topics on social media with the public attention dynamics. We have provided a novel approach to study and quantify the attention dynamics taking advantage of the real-time Hot Search List (HSL) on Weibo. The rank diversity in the top 15 ranks containing COVID-hashtags are higher than normal. This could result from the spontaneous preference from the public towards COVID-related information. More likely, it is due to an algorithmic intervention towards COVID-hashtags from the platform provider. Sina Weibo may intentionally promote the COVID-related important information to make sure people will get aware of them. In this sense, such an algorithmic intervention can be useful for the public. The empirical rank diversity could be a combined influence of both the Weibo algorithm and the spontaneous public preference. This observation shows the possibility that rank diversity could be an adequate tool to investigate further the important aspect of algorithmic intervention in social media data.

Besides exploring the attention dynamics on the Chinese social media Sina Weibo, we also studied the cumulative growth of all topics and all the COVID-topics on Twitter trending list in the United States. As is shown in the supplementary material Fig. SI1 (A), the cumulative number of all the Twitter trending topics in the United States was almost perfectly linear from January 1, 2020 to April 16, 2020. The time period that the cumulative number of all COVID-topics on Twitter trending list increases is in accordance with the rising period of the number of hashtags in the Other Countries category in Fig. \ref{fig:fig2} (B). The similarity of results on Sina Weibo HSL and Twitter trending list is a reflection that both platforms are influenced similarly by the major events worldwide during the COVID-19 pandemic. Though having more daily new topics on Twitter trending list than Weibo HSL, the number of COVID-topics on Twitter is much fewer. 
The topics on Twitter are generally shorter and have broader meaning, for example, \#QuarantineLife, while the hashtags on Weibo are more detailed, for example, \begin{CJK*}{UTF8}{gbsn} \#小区窗台演唱会庆祝解除隔离\# \end{CJK*} (\#Community window concert to celebrate the lifting of quarantine\#), contributing to the rich number of diverse hashtags on Weibo. It should be emphasized that both for Sina Weibo and Twitter the lists are produced by unknown algorithms and in the case of Sina Weibo we have been able to pinpoint direct interventions from the side of the provider into the ranking. However, the detailedness of Weibo HSL, its fixed length and the fact that HSL is the same for all users seem to make Weibo HSL more suitable to study attention dynamics through ranking than Twitter, as Twitter trending lists are without fixed length and can be personalized.


\section*{Supplementary information} 

Additional file 1. Supplementary information (PDF 2.4 MB)



\begin{backmatter}

\section*{Availability of data and materials}

The datasets supporting the conclusions of this article are available in the Attention\_Dynamics\_Sina\_Weibo\_COVID19 repository, \url{https://github.com/cuihaosabrina/Attention_Dynamics_Sina_Weibo_COVID19}

\section*{Competing interests}
The authors declare that they have no competing interests.

\section*{Funding}
JK acknowledges partial support from the H2020 project SoBigData++ (ID: 871042).

\section*{Author's contributions}

HC and JK conceived the idea and designed the study. HC carried out the data collection, HC and JK did the data analysis.
Both authors drafted the manuscript, read and approved the final manuscript. 

\section*{Acknowledgements}
Thanks are due to Márton Karsai and Tiago Peixoto for suggestions. 

\bibliographystyle{bmc-mathphys} 
\bibliography{new_bmc_article}      

\end{backmatter}

\pagebreak

\input{bmc_article_1}

\end{document}

%% file: bmc_article_1.tex
\begin{frontmatter} 

\begin{fmbox} 
\dochead{\hspace{0.2\textwidth} Supplementary information} %



\title{Attenti Attention dynamics on the Chinese social socialn media Sina Weibo during the COVID-19 pandm pandemic}


\author[
]{\inits{HC}\fnm{\hspace{0.2\textwidth}Hao} \snm{Cui}}
\author[
]{\inits{JK}\fnm{János} \snm{Kertész}}{\color{blue}{*}}


\address[id=aff2]{
	\orgname{Department of Network and Data Science, Central European University}, 
	\street{Quellenstrasse 51},                     %
	\postcode{A-1100}                                
	\city{Vienna},                              
	\cny{Austria}                                    
}

\hspace{0.2\textwidth}{\sffamily {\small {\color{blue}{*}}Correspondence: {\color{blue}{kerteszj@ceu.edu}}}}

\hspace{0.2\textwidth}{\sffamily {\small Department of Network and Data Science, Central European University, Quellenstrasse 51, A-1100, Vienna, Austria}}


\begin{artnotes}
\end{artnotes}

\end{fmbox}

\end{frontmatter}

\section*{SI1 Twitter trending COVID-topics in the United States}

Sina Weibo is the largest microblogging site in China, where Twitter, the worldwide most popular service of this kind does not operate. It is a natural idea to try to compare our observations made on Sina Weibo with Twitter attention dynamics. Unfortunately, there is no comparable statistics on Twitter to the HSL. Instead, Twitter has the service to inform about most retweeted hashtags during the last 24 hours updated on the minute basis and broken down to countries [{\color{blue}{1}}].
We have chosen to study the US tweets.

Categorization of tweets has been widely investigated [{\color{blue}{2, 3}}], 
including recent attempts to analyze the impact of COVID-related topics [{\color{blue}{4}}]
on Twitter by analyzing the sentiments to 10 words related to COVID. Twitter even created a ``COVID-19 stream" [{\color{blue}{5}}]
to promote this type of research. In spite of these, a direct comparison of our results on Sina Weibo with Twitter is hindered by a number of factors, including the different characters of the listings, the different roles hashtags play in these services and the differences due to the scripts. Nevertheless, we tried to capture at least the overall trends (see Fig. SI1). 

  \begin{figure}[!htbp]
  \includegraphics[scale=0.35]{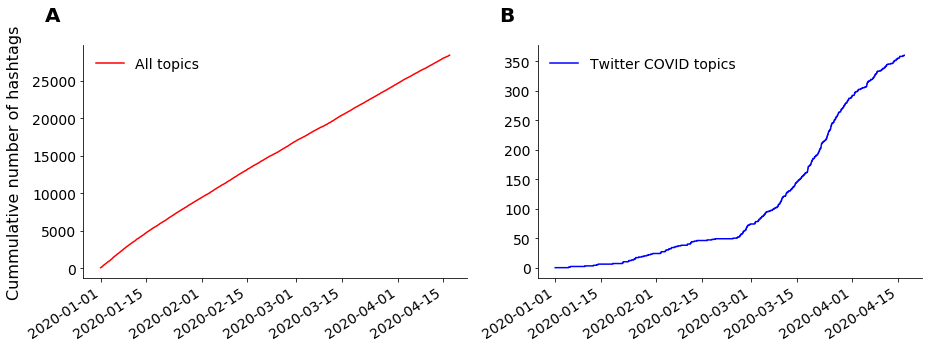}
  \renewcommand{\figurename}{Figure SI1}
  \renewcommand{\thefigure}{}
  \caption{Overview of the cumulative number of topics during the observation period on Twitter trending list in the United States from January 1, 2020 to April 16, 2020. (A) Cumulative growth of all topics. (B) Cumulative growth of COVID-related topics.}
      \end{figure}

Fig. SI1 (A) shows the cumulative number of all the Twitter trending topics in the United States is almost perfectly linear. As Fig. SI1 (B) shows, the COVID-topics on Twitter trending list first grows very slowly at the beginning phase, and then starts to increase dramatically from late February 2020. The rate of COVID-related topics is, however, much smaller in the Twitter list than on that of the Sina Weibo. 




\section*{SI2 Significance of correlations}

To understand how the categories of time series of daily new hashtags move together and whether there are blocks of categories that co-move, we presented the correlation matrices plot between the ten time series in the three periods after the outbreak. In order to get information about the significance of the correlations we apply a null model, which is created by shuffling the times of the individual values, thus smearing out the correlations. Due to the finiteness of the time series, there will be non-zero background noise level denoted by $Z$ in the null model, defining the background to which measured real correlations can be compared. $Z$ is calculated by correlating $500$ shuffled time series for each of the 10 categories. We observed that all the pairs have similar standard deviations between around 0.16 to 0.2. We take a uniform value $Z=0.2$. 

 \begin{figure}[htbp]
  \includegraphics[scale=0.43]{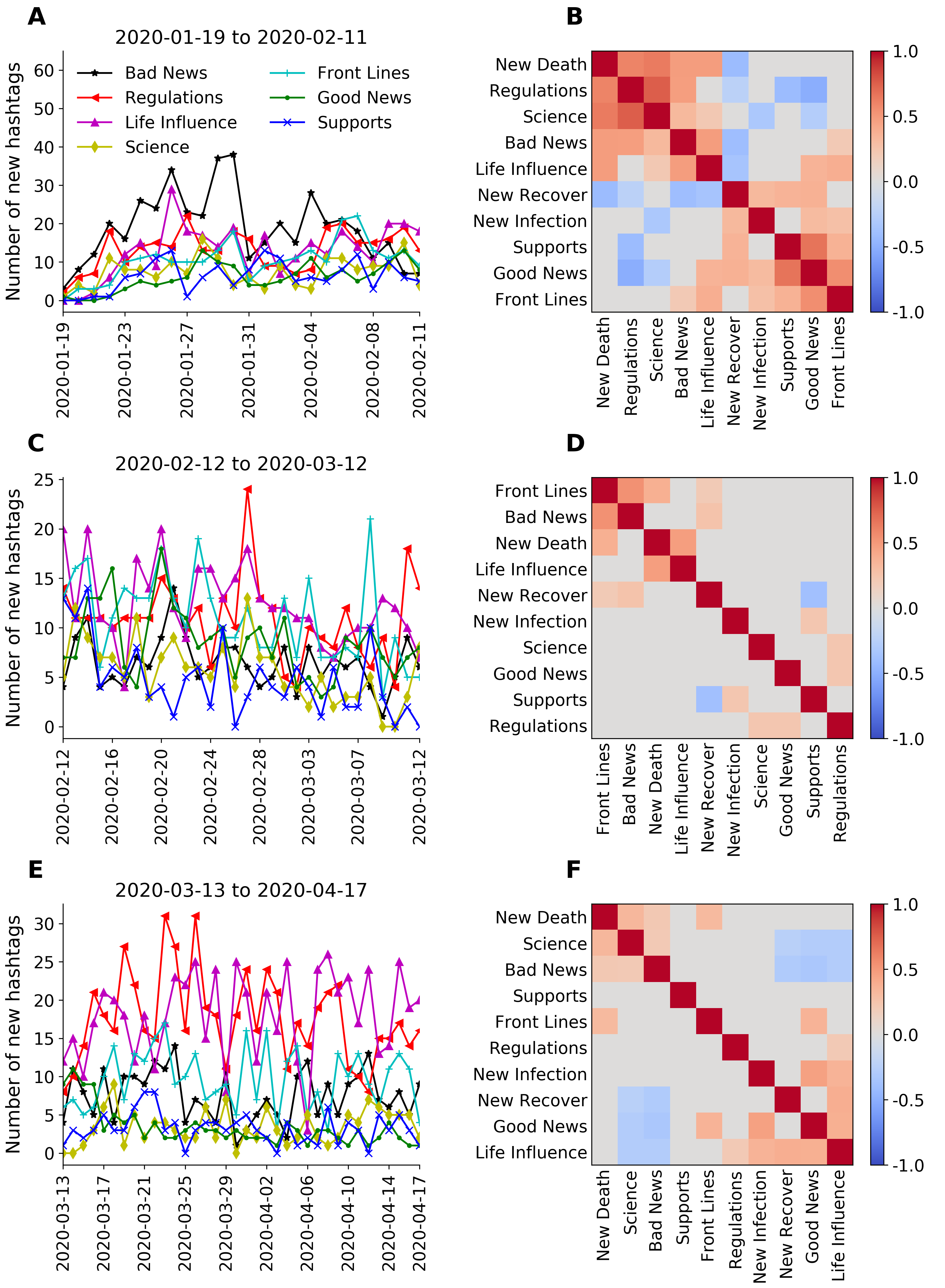}
  \renewcommand{\figurename}{Figure SI2}
  \renewcommand{\thefigure}{}
  \caption{Topical correlations of mainland China COVID-categories in three periods. Correlations lower than 0.2 are considered as insignificant and are converted to zero.}
      \end{figure}



In Fig. SI2 we show correlations where only those $C_{ij}$ correlation matrix elements are presented for which $Z<|C_{ij}|$. The figure shows the different Mainland China topical categories and their thresholded correlations in the three pandemic phases. 
In Fig. SI2 (B) most of the correlations are beyond the threshold, while in Fig. SI2 (D) very few are beyond the threshold. In Fig. SI2 (F), though some values at the upper left and lower right corners are beyond the threshold, they are much weaker than in Fig. SI2 (B).   


\section*{SI3 Categorized Sina Weibo hashtags and properties}

We showed in the main paper Fig. 4 that the gap between the top 15 ranks and the rest of the ranks in the rank diversity plot after the outbreak is caused by the COVID-hashtags. In order to further understand the properties of COVID-hashtags and how they influenced the HSL hashtag dynamics, we compared the highest rank and duration distribution of different COVID-categories with the non-COVID hashtags before and after the outbreak.

\begin{figure}[htbp]
 \begin{center}
  \includegraphics[scale=0.45]{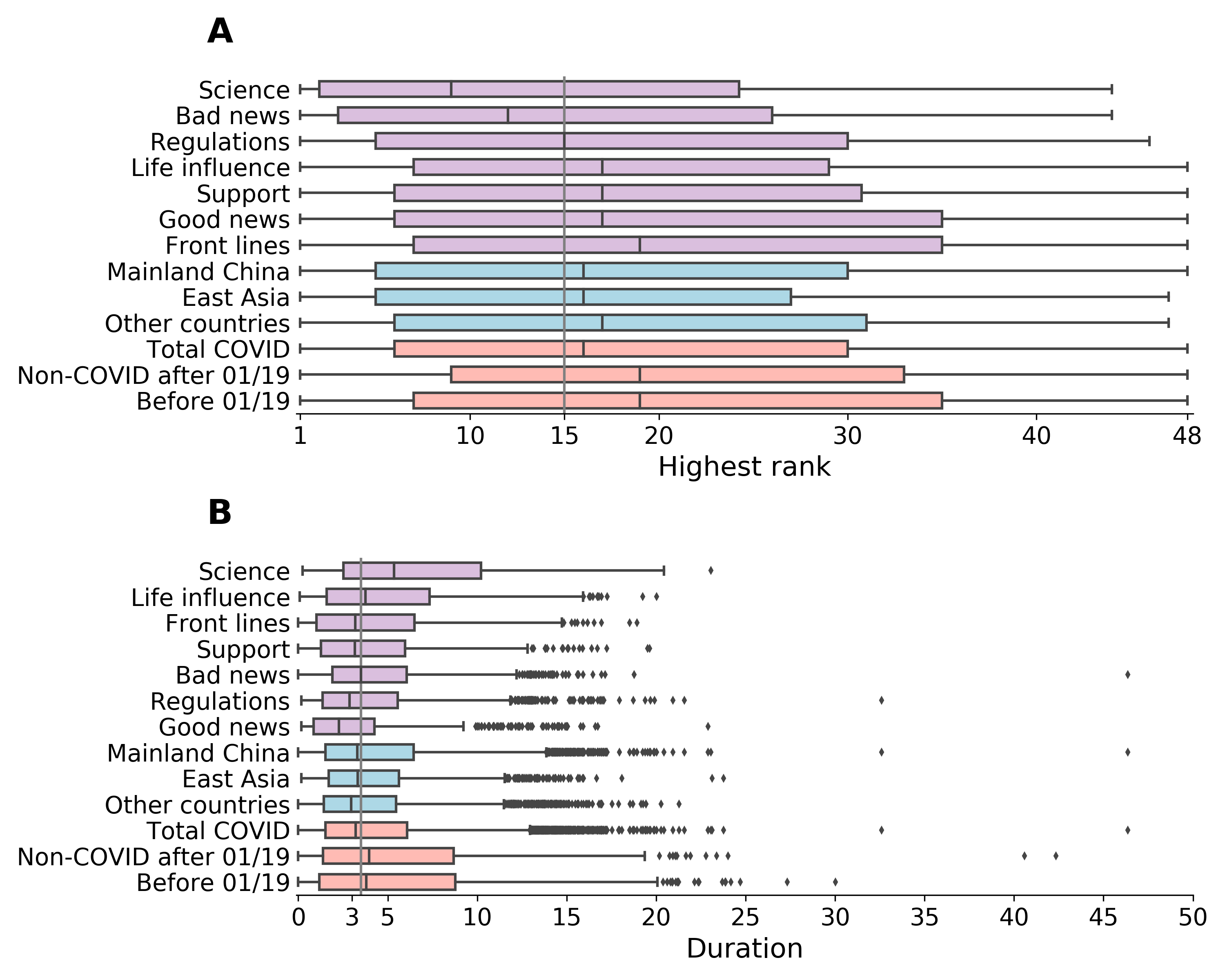}
 \end{center}
  \renewcommand{\figurename}{Figure SI3}
  \renewcommand{\thefigure}{}
  \caption{Boxplots of the highest rank (A) and duration (B) of the different categories. In both plots, the purple categories are sub-categories of Mainland China category, which is colored in blue. The blue categories are sub-categories of Total COVID category, which is colored in orange. The dots in (B) are the outlier hashtags with long duration, e.g, the
  	\#Infection Map\# in the Bad New category and the
  	\#(District Party Secretary) Will be accountable if home-confirmed patients found again in Wuhan\# in the Regulations category.
  }
\end{figure}

Fig. SI3 shows a detailed comparison of the highest rank and duration of the categorized Mainland China COVID-hashtags on Weibo Hot Search List (HSL), before and after the COVID-19 outbreak. As Fig. SI3 (A) shows, most of the categories have a median of highest rank close to 15. Science category and Bad News category are generally higher ranked than other categories. The median highest rank of the non-COVID hashtags after the outbreak is the same with that of the hashtags before the outbreak (rank 19), while the median highest rank of the COVID-hashtags is higher than both (rank 16). Fig. SI3 (B) shows the lifetime duration of the different categories. The median duration of most of the categories is less than 3.5 hours. Science category has the highest duration among all categories. Non-COVID hashtags after the outbreak (3.95 hours) and hashtags before the outbreak (3.80 hours) have similar duration distributions. The COVID-hashtags generally have shorter duration (3.21 hours) than non-COVID hashtags.


\section*{SI4 Hashtag rank trajectory examples}
      
In the main paper, we have seen strange drops in the rank diversity plot at the ranks 29 and 34 after the outbreak, this implies that the number of unique hashtags occurred at these ranks in a given time interval is smaller than usual, so that there should be hashtags staying there for unusually long time. Here we present examples of normal and abnormal hashtag rank trajectory plots, and verify there are hashtags that stay at certain ranks such as rank 29 and 34 on the HSL for a strangely long time without any fluctuation. 
      
\begin{figure}[!htbp]
  \includegraphics[scale=0.4]{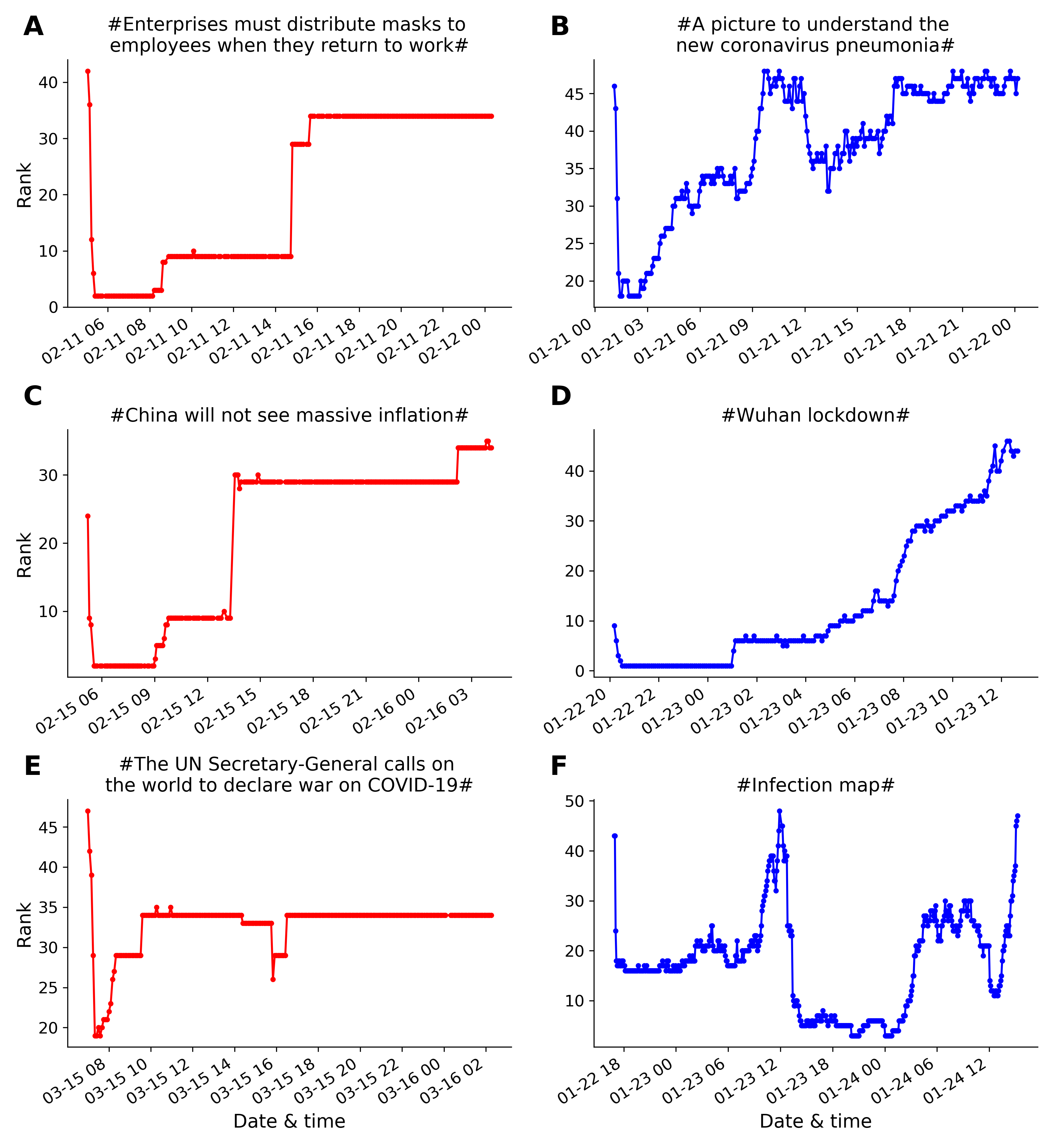}
  \renewcommand{\figurename}{Figure SI4}
  \renewcommand{\thefigure}{}
  \caption{Examples of rank trajectory plots of COVID-related hashtags. (A), (C), (E) Abnormal rank trajectory plots. (B), (D), (F) Normal rank trajectory plots.
  }
      \end{figure}

\begin{table}[!htbp] 
\renewcommand{\tablename}{Table SI1}
\renewcommand{\thetable}{}
\caption{Chinese original and translations of example hashtags in Figure SI4.}
\scalebox{0.783}{
      \begin{tabular}{ccccc}
        \hline
        Example Hashtags   & Translation\\ \hline
        \begin{CJK*}{UTF8}{gbsn} \#企业复工要为职工配发口罩\#\end{CJK*} &  \makecell{\#Enterprises must distribute masks to \\ employees when they return to work\#} \\ \hline
        
     \begin{CJK*}{UTF8}{gbsn} \#中国不会出现大规模通货膨胀\#\end{CJK*} &  \#China will not see massive inflation\# \\ \hline
         
      \begin{CJK*}{UTF8}{gbsn} \#联合国秘书长呼吁全球共同向新冠宣战\#\end{CJK*} &  \makecell{\#The UN Secretary-General calls on \\ the world to declare war on COVID-19\#} \\ \hline
      
       \begin{CJK*}{UTF8}{gbsn} \#武汉封城\#\end{CJK*} &  \#Wuhan lockdown\# \\ \hline
       
        \begin{CJK*}{UTF8}{gbsn} \#一图看懂新型冠状病毒肺炎\#\end{CJK*} &  \makecell{\#A picture to understand \\the new coronavirus pneumonia\#} \\ \hline
        
         \begin{CJK*}{UTF8}{gbsn} \#疫情地图\#\end{CJK*} &  \#Infection map\# \\ \hline

      \end{tabular} }
\end{table}

Fig. SI4 shows examples of abnormal and normal rank trajectory plots of COVID-related hashtags on Weibo HSL. In Fig. SI4 (A), (C), (E), the ranks of the hashtags stay strangely long time at ranks 29 and 34, and then disappear from the HSL. Fig. SI4 (B), (D), (F) show relatively natural fluctuations in the rank trajectory plots. The example hashtags and their translations are shown in Table SI1. The abnormal rank plots are likely due to the algorithm intervention from Sina Weibo.

%% file: new_Manuscript.bbl

\begin{thebibliography}{43}
\ifx \bisbn   \undefined \def \bisbn  #1{ISBN #1}\fi
\ifx \binits  \undefined \def \binits#1{#1}\fi
\ifx \bauthor  \undefined \def \bauthor#1{#1}\fi
\ifx \batitle  \undefined \def \batitle#1{#1}\fi
\ifx \bjtitle  \undefined \def \bjtitle#1{#1}\fi
\ifx \bvolume  \undefined \def \bvolume#1{\textbf{#1}}\fi
\ifx \byear  \undefined \def \byear#1{#1}\fi
\ifx \bissue  \undefined \def \bissue#1{#1}\fi
\ifx \bfpage  \undefined \def \bfpage#1{#1}\fi
\ifx \blpage  \undefined \def \blpage #1{#1}\fi
\ifx \burl  \undefined \def \burl#1{\textsf{#1}}\fi
\ifx \doiurl  \undefined \def \doiurl#1{\textsf{#1}}\fi
\ifx \betal  \undefined \def \betal{\textit{et al.}}\fi
\ifx \binstitute  \undefined \def \binstitute#1{#1}\fi
\ifx \binstitutionaled  \undefined \def \binstitutionaled#1{#1}\fi
\ifx \bctitle  \undefined \def \bctitle#1{#1}\fi
\ifx \beditor  \undefined \def \beditor#1{#1}\fi
\ifx \bpublisher  \undefined \def \bpublisher#1{#1}\fi
\ifx \bbtitle  \undefined \def \bbtitle#1{#1}\fi
\ifx \bedition  \undefined \def \bedition#1{#1}\fi
\ifx \bseriesno  \undefined \def \bseriesno#1{#1}\fi
\ifx \blocation  \undefined \def \blocation#1{#1}\fi
\ifx \bsertitle  \undefined \def \bsertitle#1{#1}\fi
\ifx \bsnm \undefined \def \bsnm#1{#1}\fi
\ifx \bsuffix \undefined \def \bsuffix#1{#1}\fi
\ifx \bparticle \undefined \def \bparticle#1{#1}\fi
\ifx \barticle \undefined \def \barticle#1{#1}\fi
\ifx \bconfdate \undefined \def \bconfdate #1{#1}\fi
\ifx \botherref \undefined \def \botherref #1{#1}\fi
\ifx \url \undefined \def \url#1{\textsf{#1}}\fi
\ifx \bchapter \undefined \def \bchapter#1{#1}\fi
\ifx \bbook \undefined \def \bbook#1{#1}\fi
\ifx \bcomment \undefined \def \bcomment#1{#1}\fi
\ifx \oauthor \undefined \def \oauthor#1{#1}\fi
\ifx \citeauthoryear \undefined \def \citeauthoryear#1{#1}\fi
\ifx \endbibitem  \undefined \def \endbibitem {}\fi
\ifx \bconflocation  \undefined \def \bconflocation#1{#1}\fi
\ifx \arxivurl  \undefined \def \arxivurl#1{\textsf{#1}}\fi
\csname PreBibitemsHook\endcsname

\bibitem{Wu_Huberman_2007}
\begin{barticle}
\bauthor{\bsnm{Wu}, \binits{F.}},
\bauthor{\bsnm{Huberman}, \binits{B.A.}}:
\batitle{Novelty and collective attention}.
\bjtitle{Proc. Nat. Aca. Sci.}
\bvolume{104},
\bfpage{17599}--\blpage{17601}
(\byear{2007})
\end{barticle}
\endbibitem

\bibitem{Russel_Neuman_etal_2014}
\begin{barticle}
\bauthor{\bsnm{Russell~Neuman}, \binits{W.}},
\bauthor{\bsnm{Guggenheim}, \binits{L.}},
\bauthor{\bsnm{Mo~Jang}, \binits{S.}},
\bauthor{\bsnm{Bae}, \binits{S.Y.}}:
\batitle{The dynamics of public attention: Agenda-setting theory meets big
  data}.
\bjtitle{Journal of Communication}
\bvolume{64},
\bfpage{193}--\blpage{214}
(\byear{2014})
\end{barticle}
\endbibitem

\bibitem{Twitter}
\begin{botherref}
Twitter Micoroblog and Social Network Service.
\url{https://about.twitter.com/}. Accessed December 2, 2020.
\end{botherref}
\endbibitem

\bibitem{Twitter_research}
\begin{botherref}
Twitter: Research and Experiments.
\url{https://help.twitter.com/en/rules-and-policies\#research-and-experiments}.
  Accessed December 2, 2020.
\end{botherref}
\endbibitem

\bibitem{Lehmann_etal_2012}
\begin{bchapter}
\bauthor{\bsnm{Lehmann}, \binits{J.}},
\bauthor{\bsnm{Gon\c{c}alves}, \binits{B.}},
\bauthor{\bsnm{Ramasco}, \binits{J.J.}},
\bauthor{\bsnm{Cattuto}, \binits{C.}}:
\bctitle{Dynamical Classes of Collective Attention in {Twitter}}.
In: \bbtitle{Proceedings of the 21st International Conference on World Wide Web
  (WWW)},
pp. \bfpage{251}--\blpage{260}
(\byear{2007})
\end{bchapter}
\endbibitem

\bibitem{Eom_et_al_2015}
\begin{barticle}
\bauthor{\bsnm{Eom}, \binits{Y.-H.}},
\bauthor{\bsnm{Puliga}, \binits{M.}},
\bauthor{\bsnm{Smailovi\v{c}}, \binits{J.}},
\bauthor{\bsnm{Mozeti\v{c}}, \binits{I.}},
\bauthor{\bsnm{Caldarelli}, \binits{G.}}:
\batitle{Twitter-based analysis of the dynamics of collective attention to
  political parties}.
\bjtitle{PLoS ONE}
\bvolume{10},
\bfpage{0131184}
(\byear{2015})
\end{barticle}
\endbibitem

\bibitem{Ko_etal_2014}
\begin{barticle}
\bauthor{\bsnm{Ko}, \binits{J.}},
\bauthor{\bsnm{Kwon}, \binits{H.W.}},
\bauthor{\bsnm{Kim}, \binits{H.S.}},
\bauthor{\bsnm{Lee}, \binits{K.}},
\bauthor{\bsnm{Choi}, \binits{M.Y.}}:
\batitle{Model for twitter dynamics: Public attention and time series of
  tweeting}.
\bjtitle{Physica A}
\bvolume{404},
\bfpage{141}--\blpage{149}
(\byear{2014})
\end{barticle}
\endbibitem

\bibitem{Pen_etal_2017}
\begin{barticle}
\bauthor{\bsnm{Pen}, \binits{T.-Q.}},
\bauthor{\bsnm{Sun}, \binits{G.}},
\bauthor{\bsnm{Wu}, \binits{Y.}}:
\batitle{Interplay between public attention and public emotion toward multiple
  social issues on twitter}.
\bjtitle{PLoS ONE}
\bvolume{12},
\bfpage{0167896}
(\byear{2017})
\end{barticle}
\endbibitem

\bibitem{Chew_Eysenbach_2010}
\begin{barticle}
\bauthor{\bsnm{Chew}, \binits{C.}},
\bauthor{\bsnm{Eysenbach}, \binits{G.}}:
\batitle{Pandemics in the age of twitter: Content analysis of tweets during the
  2009 h1n1 outbreak}.
\bjtitle{PLoS ONE}
\bvolume{5},
\bfpage{14118}
(\byear{2010})
\end{barticle}
\endbibitem

\bibitem{Signorini_etal_2011}
\begin{barticle}
\bauthor{\bsnm{Signorini}, \binits{A.}},
\bauthor{\bsnm{Segre}, \binits{A.M.}},
\bauthor{\bsnm{Polgreen}, \binits{P.M.}}:
\batitle{The use of twitter to track levels of disease activity and public
  concern in the {U.S.} during the influenza {A} {H1N1} pandemic}.
\bjtitle{PLoS ONE}
\bvolume{6},
\bfpage{19467}
(\byear{2011})
\end{barticle}
\endbibitem

\bibitem{van_Lent_etal_2017}
\begin{barticle}
\bauthor{\bparticle{van} \bsnm{Lent}, \binits{L.G.G.}},
\bauthor{\bsnm{Sungur}, \binits{H.}},
\bauthor{\bsnm{Kunneman}, \binits{F.A.}},
\bauthor{\bparticle{van~de} \bsnm{Velde}, \binits{B.}},
\bauthor{\bsnm{Das}, \binits{E.}}:
\batitle{Too far to care? measuring public attention and fear for ebola using
  twitter}.
\bjtitle{J. Med. Internet Res.}
\bvolume{19},
\bfpage{193}
(\byear{2017})
\end{barticle}
\endbibitem

\bibitem{Zavarrone_etal_2020}
\begin{botherref}
\oauthor{\bsnm{Zavarrone}, \binits{E.}},
\oauthor{\bsnm{Grassia}, \binits{M.G.}},
\oauthor{\bsnm{Marino}, \binits{M.}},
\oauthor{\bsnm{Cataldo}, \binits{R.}},
\oauthor{\bsnm{Mazza}, \binits{R.}},
\oauthor{\bsnm{Canestrari}, \binits{N.}}:
{CO.ME.T.A.} -- {COVID-19} media textual analysis. A dashboard for media
  monitoring.
\url{https://arxiv.org/pdf/2004.07742.pdf}. Accessed December 2, 2020.
\end{botherref}
\endbibitem

\bibitem{Lopez_etal_2020}
\begin{botherref}
\oauthor{\bsnm{Lopez}, \binits{C.E.}},
\oauthor{\bsnm{Vasu1}, \binits{M.}},
\oauthor{\bsnm{Gallemore}, \binits{C.}}:
Understanding the perception of {COVID-19} policies by mining a multilanguage
  {Twitter} dataset.
\url{https://arxiv.org/ftp/arxiv/papers/2003/2003.10359.pdf}. Accessed December
  2, 2020.
\end{botherref}
\endbibitem

\bibitem{pennycook_misinfo_2020}
\begin{barticle}
\bauthor{\bsnm{Pennycook}, \binits{G.}},
\bauthor{\bsnm{McPhetres}, \binits{J.}},
\bauthor{\bsnm{Zhang}, \binits{Y.}},
\bauthor{\bsnm{Lu}, \binits{J.G.}},
\bauthor{\bsnm{Rand}, \binits{D.G.}}:
\batitle{Fighting covid-19 misinformation on social media: Experimental
  evidence for a scalable accuracy-nudge intervention}.
\bjtitle{Psychological Science}
\bvolume{31}(\bissue{7}),
\bfpage{770}--\blpage{780}
(\byear{2020})
\end{barticle}
\endbibitem

\bibitem{mental_covid}
\begin{barticle}
\bauthor{\bsnm{Gao}, \binits{J.}},
\bauthor{\bsnm{Zheng}, \binits{P.}},
\bauthor{\bsnm{Jia}, \binits{Y.}},
\bauthor{\bsnm{Chen}, \binits{H.}},
\bauthor{\bsnm{Mao}, \binits{Y.}},
\bauthor{\bsnm{Chen}, \binits{S.}},
\bauthor{\bsnm{Wang}, \binits{Y.}},
\bauthor{\bsnm{Fu}, \binits{H.}},
\bauthor{\bsnm{Dai}, \binits{J.}}:
\batitle{Mental health problems and social media exposure during covid-19
  outbreak}.
\bjtitle{PLOS ONE}
\bvolume{15}(\bissue{4}),
\bfpage{1}--\blpage{10}
(\byear{2020})
\end{barticle}
\endbibitem

\bibitem{psy_covid}
\begin{barticle}
\bauthor{\bsnm{Dubey}, \binits{S.}},
\bauthor{\bsnm{Biswas}, \binits{P.}},
\bauthor{\bsnm{Ghosh}, \binits{R.}},
\bauthor{\bsnm{Chatterjee}, \binits{S.}},
\bauthor{\bsnm{Dubey}, \binits{M.J.}},
\bauthor{\bsnm{Chatterjee}, \binits{S.}},
\bauthor{\bsnm{Lahiri}, \binits{D.}},
\bauthor{\bsnm{Lavie}, \binits{C.J.}}:
\batitle{Psychosocial impact of covid-19}.
\bjtitle{Diabetes \& Metabolic Syndrome: Clinical Research \& Reviews}
\bvolume{14}(\bissue{5}),
\bfpage{779}--\blpage{788}
(\byear{2020})
\end{barticle}
\endbibitem

\bibitem{Weibo}
\begin{botherref}
An Introduction to {Sina} {Weibo}: {Background} and Status Quo.
\url{https://www.whatsonweibo.com/sinaweibo/}. Accessed December 2, 2020.
\end{botherref}
\endbibitem

\bibitem{Tong_Zuo_2014}
\begin{barticle}
\bauthor{\bsnm{Tong}, \binits{J.}},
\bauthor{\bsnm{Zuo}, \binits{L.}}:
\batitle{Weibo communication and government legitimacy in {China}: a
  computer-assisted analysis of {Weibo} messages on two ‘mass incidents’}.
\bjtitle{Information, Communication and Society}
\bvolume{17},
\bfpage{66}--\blpage{85}
(\byear{2014})
\end{barticle}
\endbibitem

\bibitem{Nip_Fu_2015}
\begin{barticle}
\bauthor{\bsnm{Nip}, \binits{J.Y.M.}},
\bauthor{\bsnm{Fu}, \binits{K.-w.}}:
\batitle{Networked framing between source posts and their reposts: an analysis
  of public opinion on {China's} microblogs}.
\bjtitle{Information, Communication and Society}
\bvolume{19},
\bfpage{1127}--\blpage{1149}
(\byear{2016})
\end{barticle}
\endbibitem

\bibitem{Yuner_posts}
\begin{botherref}
\oauthor{\bsnm{Y}, \binits{Z.}},
\oauthor{\bsnm{KW}, \binits{F.}},
\oauthor{\bsnm{KA}, \binits{G.}},
\oauthor{\bsnm{H}, \binits{L.}},
\oauthor{\bsnm{IC}, \binits{F.}}:
Limited early warnings and public attention to coronavirus disease 2019 in
  china, january-february, 2020: A longitudinal cohort of randomly sampled
  weibo users.
Disaster Med Public Health Prep
\textbf{1-4}
(2020)
\end{botherref}
\endbibitem

\bibitem{yuxin_descr}
\begin{barticle}
\bauthor{\bsnm{Y}, \binits{Z.}},
\bauthor{\bsnm{S}, \binits{C.}},
\bauthor{\bsnm{X}, \binits{Y.}},
\bauthor{\bsnm{H}, \binits{X.}}:
\batitle{Chinese public's attention to the covid-19 epidemic on social media:
  Observational descriptive study}.
\bjtitle{Journal of Medical Internet Research}
\bvolume{22},
\bfpage{18825}
(\byear{2020})
\end{barticle}
\endbibitem

\bibitem{xiaoya_emotion}
\begin{botherref}
\oauthor{\bsnm{Li}, \binits{X.}},
\oauthor{\bsnm{Zhou}, \binits{M.}},
\oauthor{\bsnm{Wu}, \binits{J.}},
\oauthor{\bsnm{Yuan}, \binits{A.}},
\oauthor{\bsnm{Wu}, \binits{F.}},
\oauthor{\bsnm{Li}, \binits{J.}}:
Analyzing COVID-19 on Online Social Media: Trends, Sentiments and Emotions.
\url{https://arxiv.org/pdf/2005.14464.pdf}. Accessed December 2, 2020.
\end{botherref}
\endbibitem

\bibitem{seo_corr}
\begin{barticle}
\bauthor{\bsnm{Seo}, \binits{D.-W.}},
\bauthor{\bsnm{Shin}, \binits{S.-Y.}}:
\batitle{Methods using social media and search queries to predict infectious
  disease outbreaks}.
\bjtitle{Healthc Inform Res}
\bvolume{23}(\bissue{4}),
\bfpage{343}--\blpage{348}
(\byear{2017})
\end{barticle}
\endbibitem

\bibitem{review_corr}
\begin{botherref}
\oauthor{\bsnm{Alessa}, \binits{A.}},
\oauthor{\bsnm{Faezipour}, \binits{M.}}:
A review of influenza detection and prediction through social networking sites.
Theoretical Biology and Medical Modelling
\textbf{15}(2)
(2018)
\end{botherref}
\endbibitem

\bibitem{Thomas_corr}
\begin{barticle}
\bauthor{\bsnm{Higgins}, \binits{T.S.}},
\bauthor{\bsnm{Wu}, \binits{A.W.}},
\bauthor{\bsnm{Sharma}, \binits{D.}},
\bauthor{\bsnm{Illing}, \binits{E.A.}},
\bauthor{\bsnm{Rubel}, \binits{K.}},
\bauthor{\bsnm{Ting}, \binits{J.Y.}}:
\batitle{Correlations of online search engine trends with coronavirus disease
  (covid-19) incidence: Infodemiology study}.
\bjtitle{JMIR Public Health Surveill}
\bvolume{6}(\bissue{2}),
\bfpage{19702}
(\byear{2020})
\end{barticle}
\endbibitem

\bibitem{lei_corr}
\begin{barticle}
\bauthor{\bsnm{Qin}, \binits{L.}},
\bauthor{\bsnm{Sun}, \binits{Q.}},
\bauthor{\bsnm{Wang}, \binits{Y.}},
\bauthor{\bsnm{Wu}, \binits{K.-F.}},
\bauthor{\bsnm{Chen}, \binits{M.}},
\bauthor{\bsnm{Shia}, \binits{B.-C.}},
\bauthor{\bsnm{Wu}, \binits{S.-Y.}}:
\batitle{Prediction of number of cases of 2019 novel coronavirus (covid-19)
  using social media search index}.
\bjtitle{Int J Environ Res Public Health}
\bvolume{17}(\bissue{7}),
\bfpage{2365}
(\byear{2020})
\end{barticle}
\endbibitem

\bibitem{cuilian_covidcorr}
\begin{botherref}
\oauthor{\bsnm{Li}, \binits{C.}},
\oauthor{\bsnm{Chen}, \binits{L.J.}},
\oauthor{\bsnm{Chen}, \binits{X.}},
\oauthor{\bsnm{Zhang}, \binits{M.}},
\oauthor{\bsnm{Pang}, \binits{C.P.}},
\oauthor{\bsnm{Chen}, \binits{H.}}:
Retrospective analysis of the possibility of predicting the covid-19 outbreak
  from internet searches and social media data, china, 2020.
Euro Surveill
\textbf{25}(10)
(2020)
\end{botherref}
\endbibitem

\bibitem{cholera}
\begin{barticle}
\bauthor{\bsnm{Chunara}, \binits{R.}},
\bauthor{\bsnm{Andrews}, \binits{J.R.}},
\bauthor{\bsnm{Brownstein}, \binits{J.S.}}:
\batitle{Social and news media enable estimation of epidemiological patterns
  early in the 2010 haitian cholera outbreak}.
\bjtitle{The American Society of Tropical Medicine and Hygiene}
\bvolume{86}(\bissue{1}),
\bfpage{39}--\blpage{45}
(\byear{2012})
\end{barticle}
\endbibitem

\bibitem{kui_ebola}
\begin{barticle}
\bauthor{\bsnm{Liu}, \binits{K.}},
\bauthor{\bsnm{Li}, \binits{L.}},
\bauthor{\bsnm{Jiang}, \binits{T.}},
\bauthor{\bsnm{Chen}, \binits{B.}},
\bauthor{\bsnm{Jiang}, \binits{Z.}},
\bauthor{\bsnm{Wang}, \binits{Z.}},
\bauthor{\bsnm{Chen}, \binits{Y.}},
\bauthor{\bsnm{Jiang}, \binits{J.}},
\bauthor{\bsnm{Gu}, \binits{H.}}:
\batitle{Chinese public attention to the outbreak of ebola in west africa:
  Evidence from the online big data platform}.
\bjtitle{Int J Environ Res Public Health}
\bvolume{13}(\bissue{8}),
\bfpage{780}
(\byear{2016})
\end{barticle}
\endbibitem

\bibitem{Blumm_etal_2012}
\begin{barticle}
\bauthor{\bsnm{Blumm}, \binits{N.}},
\bauthor{\bsnm{Ghoshal}, \binits{G.}},
\bauthor{\bsnm{Forr\'o}, \binits{Z.}},
\bauthor{\bsnm{Schich}, \binits{M.}},
\bauthor{\bsnm{Bianconi}, \binits{G.}},
\bauthor{\bsnm{Bouchaud}, \binits{J.-P.}},
\bauthor{\bsnm{Barab\'asi}, \binits{A.-L.}}:
\batitle{Dynamics of ranking processes in complex systems}.
\bjtitle{Physical Review Letters}
\bvolume{109},
\bfpage{128701}
(\byear{2012})
\end{barticle}
\endbibitem

\bibitem{Criado_2013}
\begin{barticle}
\bauthor{\bsnm{Criado}, \binits{R.}},
\bauthor{\bsnm{Garcia}, \binits{E.}},
\bauthor{\bsnm{Pedroche}, \binits{F.}},
\bauthor{\bsnm{Romance}, \binits{M.}}:
\batitle{A new method for comparing rankings through complex networks: Model
  and analysis of competitiveness of major european soccer leagues}.
\bjtitle{Chaos}
\bvolume{23},
\bfpage{043114}
(\byear{2013})
\end{barticle}
\endbibitem

\bibitem{Morales_etal_2016}
\begin{barticle}
\bauthor{\bsnm{Morales}, \binits{J.A.}},
\bauthor{\bsnm{Sánchez}, \binits{S.}},
\bauthor{\bsnm{Flores}, \binits{J.}},
\bauthor{\bsnm{Pineda}, \binits{C.}},
\bauthor{\bsnm{Gershenson}, \binits{C.}},
\bauthor{\bsnm{Cocho}, \binits{G.}},
\bauthor{\bsnm{Zizumbo}, \binits{J.}},
\bauthor{\bsnm{Rodr\'iguez}, \binits{R.F.}},
\bauthor{\bsnm{I$\mathrm{\tilde{n}}$iguez}, \binits{G.}}:
\batitle{Generic temporal features of performance rankings in sports and
  games}.
\bjtitle{EPJ Data Science}
\bvolume{5},
\bfpage{33}
(\byear{2016})
\end{barticle}
\endbibitem

\bibitem{stable_rank}
\begin{barticle}
\bauthor{\bsnm{Morales}, \binits{J.A.}},
\bauthor{\bsnm{Colman}, \binits{E.}},
\bauthor{\bsnm{Sánchez}, \binits{S.}},
\bauthor{\bsnm{Sánchez-Puig}, \binits{F.}},
\bauthor{\bsnm{Pineda}, \binits{C.}},
\bauthor{\bsnm{Iñiguez}, \binits{G.}},
\bauthor{\bsnm{Cocho}, \binits{G.}},
\bauthor{\bsnm{Flores}, \binits{J.}},
\bauthor{\bsnm{Gershenson}, \binits{C.}}:
\batitle{Rank dynamics of word usage at multiple scales}.
\bjtitle{Frontiers in Physics}
\bvolume{6},
\bfpage{45}
(\byear{2018})
\end{barticle}
\endbibitem

\bibitem{alshaabi2020worlds}
\begin{botherref}
\oauthor{\bsnm{Alshaabi}, \binits{T.}},
\oauthor{\bsnm{Minot}, \binits{J.R.}},
\oauthor{\bsnm{Arnold}, \binits{M.V.}},
\oauthor{\bsnm{Adams}, \binits{J.L.}},
\oauthor{\bsnm{Dewhurst}, \binits{D.R.}},
\oauthor{\bsnm{Reagan}, \binits{A.J.}},
\oauthor{\bsnm{Muhamad}, \binits{R.}},
\oauthor{\bsnm{Danforth}, \binits{C.M.}},
\oauthor{\bsnm{Dodds}, \binits{P.S.}}:
How the world's collective attention is being paid to a pandemic: COVID-19
  related n-gram time series for 24 languages on Twitter.
\url{https://arxiv.org/pdf/2003.12614.pdf}. Accessed December 2, 2020.
\end{botherref}
\endbibitem

\bibitem{dewhurst2020divergent}
\begin{botherref}
\oauthor{\bsnm{Dewhurst}, \binits{D.R.}},
\oauthor{\bsnm{Alshaabi}, \binits{T.}},
\oauthor{\bsnm{Arnold}, \binits{M.V.}},
\oauthor{\bsnm{Minot}, \binits{J.R.}},
\oauthor{\bsnm{Danforth}, \binits{C.M.}},
\oauthor{\bsnm{Dodds}, \binits{P.S.}}:
Divergent modes of online collective attention to the COVID-19 pandemic are
  associated with future caseload variance.
\url{https://arxiv.org/pdf/2004.03516.pdf}. Accessed December 2, 2020.
\end{botherref}
\endbibitem

\bibitem{Weibomau}
\begin{botherref}
Weibo Reports First Quarter 2020 Unaudited Financial Results.
\url{http://ir.weibo.com/news-releases/news-release-details/weibo-reports-first-quarter-2020-unaudited-financial-results/}.
  Accessed December 2, 2020.
\end{botherref}
\endbibitem

\bibitem{Weibointro}
\begin{botherref}
\oauthor{\bsnm{Wang}, \binits{Y.}}:
An {Introduction} to {Sina} {Weibo} for {Journalists}.
\url{https://www.interhacktives.com/2018/02/22/how-to-use-sina-weibo-as-a-journalist/}.
  Accessed December 2, 2020.
\end{botherref}
\endbibitem

\bibitem{Weiboindex}
\begin{botherref}
\oauthor{\bsnm{Service}, \binits{W.C.}}:
Common Questions on the Rules of Real-time Hot-Search-List, Hot-Message-List
  and Hot-Topic-List.
\url{https://www.weibo.com/ttarticle/p/show?id=2309404007731978739654}.
  Accessed December 2, 2020.
\end{botherref}
\endbibitem

\bibitem{Weiboads}
\begin{botherref}
Weibo {Advertising}.
\url{https://www.marketingtochina.com/weibo-advertising/}. Accessed December 2,
  2020.
\end{botherref}
\endbibitem

\bibitem{nhs}
\begin{botherref}
National Health Commission of People's Republic of China.
\url{http://www.nhc.gov.cn/xcs/xxgzbd/gzbd_index.shtml}. Accessed December 2,
  2020.
\end{botherref}
\endbibitem

\bibitem{Feb12}
\begin{botherref}
{China confirms 15152 new coronavirus cases, 254 additional deaths}.
\url{https://www.cnbc.com/2020/02/13/coronavirus-latest-updates-china-hubei.html}.
  Accessed December 2, 2020.
\end{botherref}
\endbibitem

\bibitem{importcase}
\begin{botherref}
\oauthor{\bsnm{Sajid}, \binits{I.}}:
China reports 99 new virus cases, majority imported.
\url{https://www.aa.com.tr/en/asia-pacific/china-reports-99-new-virus-cases-majority-imported/1801667}.
  Accessed December 2, 2020.
\end{botherref}
\endbibitem

\bibitem{smoothen}
\begin{botherref}
Savitzky–Golay filter.
\url{https://en.wikipedia.org/wiki/Savitzky-Golay_filter}. Accessed December 2,
  2020.
\end{botherref}
\endbibitem

\end{thebibliography}

\newcommand{\BMCxmlcomment}[1]{}

\BMCxmlcomment{

<refgrp>

<bibl id="B1">
  <title><p>Novelty and collective attention</p></title>
  <aug>
    <au><snm>Wu</snm><fnm>F</fnm></au>
    <au><snm>Huberman</snm><fnm>BA</fnm></au>
  </aug>
  <source>Proc. Nat. Aca. Sci.</source>
  <pubdate>2007</pubdate>
  <volume>104</volume>
  <fpage>17599</fpage>
  <lpage>17601</lpage>
</bibl>

<bibl id="B2">
  <title><p>The Dynamics of Public Attention: Agenda-Setting Theory Meets Big
  Data</p></title>
  <aug>
    <au><snm>Russell Neuman</snm><fnm>W.</fnm></au>
    <au><snm>Guggenheim</snm><fnm>L.</fnm></au>
    <au><snm>Mo Jang</snm><fnm>S.</fnm></au>
    <au><snm>Bae</snm><fnm>S. Y.</fnm></au>
  </aug>
  <source>Journal of Communication</source>
  <pubdate>2014</pubdate>
  <volume>64</volume>
  <fpage>193</fpage>
  <lpage>214</lpage>
</bibl>

<bibl id="B3">
  <title><p>Twitter Micoroblog and Social Network Service</p></title>
  <source>\url{https://about.twitter.com/}. Accessed December 2, 2020.</source>
</bibl>

<bibl id="B4">
  <title><p>Twitter: Research and Experiments</p></title>
  <source>\url{https://help.twitter.com/en/rules-and-policies\#research-and-experiments}.
  Accessed December 2, 2020.</source>
</bibl>

<bibl id="B5">
  <title><p>Dynamical classes of collective attention in {Twitter}</p></title>
  <aug>
    <au><snm>Lehmann</snm><fnm>J.</fnm></au>
    <au><snm>Gon\c{c}alves</snm><fnm>B.</fnm></au>
    <au><snm>Ramasco</snm><fnm>J.J.</fnm></au>
    <au><snm>Cattuto</snm><fnm>C.</fnm></au>
  </aug>
  <source>Proceedings of the 21st international conference on World Wide Web
  (WWW)</source>
  <pubdate>2007</pubdate>
  <fpage>251</fpage>
  <lpage>260</lpage>
</bibl>

<bibl id="B6">
  <title><p>Twitter-Based Analysis of the Dynamics of Collective Attention to
  Political Parties</p></title>
  <aug>
    <au><snm>Eom</snm><fnm>YH</fnm></au>
    <au><snm>Puliga</snm><fnm>M</fnm></au>
    <au><snm>Smailovi\v{c}</snm><fnm>J</fnm></au>
    <au><snm>Mozeti\v{c}</snm><fnm>I</fnm></au>
    <au><snm>Caldarelli</snm><fnm>G</fnm></au>
  </aug>
  <source>PLoS ONE</source>
  <pubdate>2015</pubdate>
  <volume>10</volume>
  <fpage>e0131184</fpage>
</bibl>

<bibl id="B7">
  <title><p>Model for Twitter dynamics: Public attention and time series of
  tweeting</p></title>
  <aug>
    <au><snm>Ko</snm><fnm>J.</fnm></au>
    <au><snm>Kwon</snm><fnm>H.W.</fnm></au>
    <au><snm>Kim</snm><fnm>H.S.</fnm></au>
    <au><snm>Lee</snm><fnm>K.</fnm></au>
    <au><snm>Choi</snm><fnm>M.Y.</fnm></au>
  </aug>
  <source>Physica A</source>
  <pubdate>2014</pubdate>
  <volume>404</volume>
  <fpage>141</fpage>
  <lpage>149</lpage>
</bibl>

<bibl id="B8">
  <title><p>Interplay between Public Attention and Public Emotion toward
  Multiple Social Issues on Twitter</p></title>
  <aug>
    <au><snm>Pen</snm><fnm>TQ</fnm></au>
    <au><snm>Sun</snm><fnm>G</fnm></au>
    <au><snm>Wu</snm><fnm>Y</fnm></au>
  </aug>
  <source>PLoS ONE</source>
  <pubdate>2017</pubdate>
  <volume>12</volume>
  <fpage>e0167896</fpage>
</bibl>

<bibl id="B9">
  <title><p>Pandemics in the Age of Twitter: Content Analysis of Tweets during
  the 2009 H1N1 Outbreak</p></title>
  <aug>
    <au><snm>Chew</snm><fnm>C.</fnm></au>
    <au><snm>Eysenbach</snm><fnm>G.</fnm></au>
  </aug>
  <source>PLoS ONE</source>
  <pubdate>2010</pubdate>
  <volume>5</volume>
  <fpage>e14118</fpage>
</bibl>

<bibl id="B10">
  <title><p>The Use of Twitter to Track Levels of Disease Activity and Public
  Concern in the {U.S.} during the Influenza {A} {H1N1} Pandemic</p></title>
  <aug>
    <au><snm>Signorini</snm><fnm>A</fnm></au>
    <au><snm>Segre</snm><fnm>AM</fnm></au>
    <au><snm>Polgreen</snm><fnm>PM</fnm></au>
  </aug>
  <source>PLoS ONE</source>
  <pubdate>2011</pubdate>
  <volume>6</volume>
  <fpage>e19467</fpage>
</bibl>

<bibl id="B11">
  <title><p>Too Far to Care? Measuring Public Attention and Fear for Ebola
  Using Twitter</p></title>
  <aug>
    <au><snm>Lent</snm><fnm>LGG</fnm></au>
    <au><snm>Sungur</snm><fnm>H</fnm></au>
    <au><snm>Kunneman</snm><fnm>FA</fnm></au>
    <au><snm>Velde</snm><fnm>B</fnm></au>
    <au><snm>Das</snm><fnm>E</fnm></au>
  </aug>
  <source>J. Med. Internet Res.</source>
  <pubdate>2017</pubdate>
  <volume>19</volume>
  <fpage>e193</fpage>
</bibl>

<bibl id="B12">
  <title><p>{CO.ME.T.A.} -- {COVID-19} media textual analysis. A dashboard for
  media monitoring</p></title>
  <aug>
    <au><snm>Zavarrone</snm><fnm>E</fnm></au>
    <au><snm>Grassia</snm><fnm>MG</fnm></au>
    <au><snm>Marino</snm><fnm>M</fnm></au>
    <au><snm>Cataldo</snm><fnm>R</fnm></au>
    <au><snm>Mazza</snm><fnm>R</fnm></au>
    <au><snm>Canestrari</snm><fnm>N</fnm></au>
  </aug>
  <source>\url{https://arxiv.org/pdf/2004.07742.pdf}. Accessed December 2,
  2020.</source>
</bibl>

<bibl id="B13">
  <title><p>Understanding the perception of {COVID-19} policies by mining a
  multilanguage {Twitter} dataset</p></title>
  <aug>
    <au><snm>Lopez</snm><fnm>CE</fnm></au>
    <au><snm>Vasu1</snm><fnm>M</fnm></au>
    <au><snm>Gallemore</snm><fnm>C</fnm></au>
  </aug>
  <source>\url{https://arxiv.org/ftp/arxiv/papers/2003/2003.10359.pdf}.
  Accessed December 2, 2020.</source>
</bibl>

<bibl id="B14">
  <title><p>Fighting COVID-19 Misinformation on Social Media: Experimental
  Evidence for a Scalable Accuracy-Nudge Intervention</p></title>
  <aug>
    <au><snm>Pennycook</snm><fnm>G</fnm></au>
    <au><snm>McPhetres</snm><fnm>J</fnm></au>
    <au><snm>Zhang</snm><fnm>Y</fnm></au>
    <au><snm>Lu</snm><fnm>JG</fnm></au>
    <au><snm>Rand</snm><fnm>DG</fnm></au>
  </aug>
  <source>Psychological Science</source>
  <pubdate>2020</pubdate>
  <volume>31</volume>
  <issue>7</issue>
  <fpage>770</fpage>
  <lpage>780</lpage>
</bibl>

<bibl id="B15">
  <title><p>Mental health problems and social media exposure during COVID-19
  outbreak</p></title>
  <aug>
    <au><snm>Gao</snm><fnm>J</fnm></au>
    <au><snm>Zheng</snm><fnm>P</fnm></au>
    <au><snm>Jia</snm><fnm>Y</fnm></au>
    <au><snm>Chen</snm><fnm>H</fnm></au>
    <au><snm>Mao</snm><fnm>Y</fnm></au>
    <au><snm>Chen</snm><fnm>S</fnm></au>
    <au><snm>Wang</snm><fnm>Y</fnm></au>
    <au><snm>Fu</snm><fnm>H</fnm></au>
    <au><snm>Dai</snm><fnm>J</fnm></au>
  </aug>
  <source>PLOS ONE</source>
  <publisher>Public Library of Science</publisher>
  <pubdate>2020</pubdate>
  <volume>15</volume>
  <issue>4</issue>
  <fpage>1</fpage>
  <lpage>10</lpage>
</bibl>

<bibl id="B16">
  <title><p>Psychosocial impact of COVID-19</p></title>
  <aug>
    <au><snm>Dubey</snm><fnm>S</fnm></au>
    <au><snm>Biswas</snm><fnm>P</fnm></au>
    <au><snm>Ghosh</snm><fnm>R</fnm></au>
    <au><snm>Chatterjee</snm><fnm>S</fnm></au>
    <au><snm>Dubey</snm><fnm>MJ</fnm></au>
    <au><snm>Chatterjee</snm><fnm>S</fnm></au>
    <au><snm>Lahiri</snm><fnm>D</fnm></au>
    <au><snm>Lavie</snm><fnm>CJ</fnm></au>
  </aug>
  <source>Diabetes \& Metabolic Syndrome: Clinical Research \& Reviews</source>
  <pubdate>2020</pubdate>
  <volume>14</volume>
  <issue>5</issue>
  <fpage>779</fpage>
  <lpage>788</lpage>
  <url>http://www.sciencedirect.com/science/article/pii/S1871402120301545</url>
</bibl>

<bibl id="B17">
  <title><p>An Introduction to {Sina} {Weibo}: {Background} and Status
  Quo</p></title>
  <source>\url{https://www.whatsonweibo.com/sinaweibo/}. Accessed December 2,
  2020.</source>
</bibl>

<bibl id="B18">
  <title><p>Weibo communication and government legitimacy in {China}: a
  computer-assisted analysis of {Weibo} messages on two ‘mass
  incidents’</p></title>
  <aug>
    <au><snm>Tong</snm><fnm>J</fnm></au>
    <au><snm>Zuo</snm><fnm>L</fnm></au>
  </aug>
  <source>Information, Communication and Society</source>
  <pubdate>2014</pubdate>
  <volume>17</volume>
  <fpage>66</fpage>
  <lpage>85</lpage>
</bibl>

<bibl id="B19">
  <title><p>Networked framing between source posts and their reposts: an
  analysis of public opinion on {China's} microblogs</p></title>
  <aug>
    <au><snm>Nip</snm><fnm>JYM</fnm></au>
    <au><snm>Fu</snm><fnm>K</fnm></au>
  </aug>
  <source>Information, Communication and Society</source>
  <pubdate>2016</pubdate>
  <volume>19</volume>
  <fpage>1127</fpage>
  <lpage>1149</lpage>
</bibl>

<bibl id="B20">
  <title><p>Limited Early Warnings and Public Attention to Coronavirus Disease
  2019 in China, January-February, 2020: A Longitudinal Cohort of Randomly
  Sampled Weibo Users</p></title>
  <aug>
    <au><snm>Y</snm><fnm>Z</fnm></au>
    <au><snm>KW</snm><fnm>F</fnm></au>
    <au><snm>KA</snm><fnm>G</fnm></au>
    <au><snm>H</snm><fnm>L</fnm></au>
    <au><snm>IC</snm><fnm>F</fnm></au>
  </aug>
  <source>Disaster Med Public Health Prep</source>
  <pubdate>2020</pubdate>
  <volume>1-4</volume>
</bibl>

<bibl id="B21">
  <title><p>Chinese Public's Attention to the COVID-19 Epidemic on Social
  Media: Observational Descriptive Study</p></title>
  <aug>
    <au><snm>Y</snm><fnm>Z</fnm></au>
    <au><snm>S</snm><fnm>C</fnm></au>
    <au><snm>X</snm><fnm>Y</fnm></au>
    <au><snm>H</snm><fnm>X</fnm></au>
  </aug>
  <source>Journal of Medical Internet Research</source>
  <pubdate>2020</pubdate>
  <volume>22</volume>
  <fpage>e18825</fpage>
</bibl>

<bibl id="B22">
  <title><p>Analyzing COVID-19 on Online Social Media: Trends, Sentiments and
  Emotions</p></title>
  <aug>
    <au><snm>Li</snm><fnm>X</fnm></au>
    <au><snm>Zhou</snm><fnm>M</fnm></au>
    <au><snm>Wu</snm><fnm>J</fnm></au>
    <au><snm>Yuan</snm><fnm>A</fnm></au>
    <au><snm>Wu</snm><fnm>F</fnm></au>
    <au><snm>Li</snm><fnm>J</fnm></au>
  </aug>
  <source>\url{https://arxiv.org/pdf/2005.14464.pdf}. Accessed December 2,
  2020.</source>
</bibl>

<bibl id="B23">
  <title><p>Methods Using Social Media and Search Queries to Predict Infectious
  Disease Outbreaks</p></title>
  <aug>
    <au><snm>Seo</snm><fnm>DW</fnm></au>
    <au><snm>Shin</snm><fnm>SY</fnm></au>
  </aug>
  <source>Healthc Inform Res</source>
  <pubdate>2017</pubdate>
  <volume>23</volume>
  <issue>4</issue>
  <fpage>343</fpage>
  <lpage>348</lpage>
</bibl>

<bibl id="B24">
  <title><p>A review of influenza detection and prediction through social
  networking sites</p></title>
  <aug>
    <au><snm>Alessa</snm><fnm>A</fnm></au>
    <au><snm>Faezipour</snm><fnm>M</fnm></au>
  </aug>
  <source>Theoretical Biology and Medical Modelling</source>
  <pubdate>2018</pubdate>
  <volume>15</volume>
  <issue>2</issue>
</bibl>

<bibl id="B25">
  <title><p>Correlations of Online Search Engine Trends With Coronavirus
  Disease (COVID-19) Incidence: Infodemiology Study</p></title>
  <aug>
    <au><snm>Higgins</snm><fnm>TS</fnm></au>
    <au><snm>Wu</snm><fnm>AW</fnm></au>
    <au><snm>Sharma</snm><fnm>D</fnm></au>
    <au><snm>Illing</snm><fnm>EA</fnm></au>
    <au><snm>Rubel</snm><fnm>K</fnm></au>
    <au><snm>Ting</snm><fnm>JY</fnm></au>
  </aug>
  <source>JMIR Public Health Surveill</source>
  <pubdate>2020</pubdate>
  <volume>6</volume>
  <issue>2</issue>
  <fpage>e19702</fpage>
</bibl>

<bibl id="B26">
  <title><p>Prediction of Number of Cases of 2019 Novel Coronavirus (COVID-19)
  Using Social Media Search Index</p></title>
  <aug>
    <au><snm>Qin</snm><fnm>L</fnm></au>
    <au><snm>Sun</snm><fnm>Q</fnm></au>
    <au><snm>Wang</snm><fnm>Y</fnm></au>
    <au><snm>Wu</snm><fnm>KF</fnm></au>
    <au><snm>Chen</snm><fnm>M</fnm></au>
    <au><snm>Shia</snm><fnm>BC</fnm></au>
    <au><snm>Wu</snm><fnm>SY</fnm></au>
  </aug>
  <source>Int J Environ Res Public Health</source>
  <pubdate>2020</pubdate>
  <volume>17</volume>
  <issue>7</issue>
  <fpage>2365</fpage>
</bibl>

<bibl id="B27">
  <title><p>Retrospective analysis of the possibility of predicting the
  COVID-19 outbreak from Internet searches and social media data, China,
  2020</p></title>
  <aug>
    <au><snm>Li</snm><fnm>C</fnm></au>
    <au><snm>Chen</snm><fnm>LJ</fnm></au>
    <au><snm>Chen</snm><fnm>X</fnm></au>
    <au><snm>Zhang</snm><fnm>M</fnm></au>
    <au><snm>Pang</snm><fnm>CP</fnm></au>
    <au><snm>Chen</snm><fnm>H</fnm></au>
  </aug>
  <source>Euro Surveill</source>
  <pubdate>2020</pubdate>
  <volume>25</volume>
  <issue>10</issue>
</bibl>

<bibl id="B28">
  <title><p>Social and News Media Enable Estimation of Epidemiological Patterns
  Early in the 2010 Haitian Cholera Outbreak</p></title>
  <aug>
    <au><snm>Chunara</snm><fnm>R</fnm></au>
    <au><snm>Andrews</snm><fnm>JR</fnm></au>
    <au><snm>Brownstein</snm><fnm>JS</fnm></au>
  </aug>
  <source>The American Society of Tropical Medicine and Hygiene</source>
  <pubdate>2012</pubdate>
  <volume>86</volume>
  <issue>1</issue>
  <fpage>39</fpage>
  <lpage>45</lpage>
</bibl>

<bibl id="B29">
  <title><p>Chinese Public Attention to the Outbreak of Ebola in West Africa:
  Evidence from the Online Big Data Platform</p></title>
  <aug>
    <au><snm>Liu</snm><fnm>K</fnm></au>
    <au><snm>Li</snm><fnm>L</fnm></au>
    <au><snm>Jiang</snm><fnm>T</fnm></au>
    <au><snm>Chen</snm><fnm>B</fnm></au>
    <au><snm>Jiang</snm><fnm>Z</fnm></au>
    <au><snm>Wang</snm><fnm>Z</fnm></au>
    <au><snm>Chen</snm><fnm>Y</fnm></au>
    <au><snm>Jiang</snm><fnm>J</fnm></au>
    <au><snm>Gu</snm><fnm>H</fnm></au>
  </aug>
  <source>Int J Environ Res Public Health</source>
  <pubdate>2016</pubdate>
  <volume>13</volume>
  <issue>8</issue>
  <fpage>780</fpage>
</bibl>

<bibl id="B30">
  <title><p>Dynamics of Ranking Processes in Complex Systems</p></title>
  <aug>
    <au><snm>Blumm</snm><fnm>N</fnm></au>
    <au><snm>Ghoshal</snm><fnm>G</fnm></au>
    <au><snm>Forr\'o</snm><fnm>Z</fnm></au>
    <au><snm>Schich</snm><fnm>M</fnm></au>
    <au><snm>Bianconi</snm><fnm>G</fnm></au>
    <au><snm>Bouchaud</snm><fnm>JP</fnm></au>
    <au><snm>Barab\'asi</snm><fnm>AL</fnm></au>
  </aug>
  <source>Physical Review Letters</source>
  <pubdate>2012</pubdate>
  <volume>109</volume>
  <fpage>128701</fpage>
</bibl>

<bibl id="B31">
  <title><p>A new method for comparing rankings through complex networks: Model
  and analysis of competitiveness of major European soccer leagues</p></title>
  <aug>
    <au><snm>Criado</snm><fnm>R.</fnm></au>
    <au><snm>Garcia</snm><fnm>E.</fnm></au>
    <au><snm>Pedroche</snm><fnm>F.</fnm></au>
    <au><snm>Romance</snm><fnm>M.</fnm></au>
  </aug>
  <source>Chaos</source>
  <pubdate>2013</pubdate>
  <volume>23</volume>
  <fpage>043114</fpage>
</bibl>

<bibl id="B32">
  <title><p>Generic temporal features of performance rankings in sports and
  games</p></title>
  <aug>
    <au><snm>Morales</snm><fnm>JA</fnm></au>
    <au><snm>Sánchez</snm><fnm>S</fnm></au>
    <au><snm>Flores</snm><fnm>J</fnm></au>
    <au><snm>Pineda</snm><fnm>C</fnm></au>
    <au><snm>Gershenson</snm><fnm>C</fnm></au>
    <au><snm>Cocho</snm><fnm>G</fnm></au>
    <au><snm>Zizumbo</snm><fnm>J</fnm></au>
    <au><snm>Rodr\'iguez</snm><fnm>RF</fnm></au>
    <au><snm>I$\mathrm{\tilde{n}}$iguez</snm><fnm>G</fnm></au>
  </aug>
  <source>EPJ Data Science</source>
  <pubdate>2016</pubdate>
  <volume>5</volume>
  <fpage>33</fpage>
</bibl>

<bibl id="B33">
  <title><p>Rank Dynamics of Word Usage at Multiple Scales</p></title>
  <aug>
    <au><snm>Morales</snm><fnm>JA</fnm></au>
    <au><snm>Colman</snm><fnm>E</fnm></au>
    <au><snm>Sánchez</snm><fnm>S</fnm></au>
    <au><snm>Sánchez Puig</snm><fnm>F</fnm></au>
    <au><snm>Pineda</snm><fnm>C</fnm></au>
    <au><snm>Iñiguez</snm><fnm>G</fnm></au>
    <au><snm>Cocho</snm><fnm>G</fnm></au>
    <au><snm>Flores</snm><fnm>J</fnm></au>
    <au><snm>Gershenson</snm><fnm>C</fnm></au>
  </aug>
  <source>Frontiers in Physics</source>
  <pubdate>2018</pubdate>
  <volume>6</volume>
  <fpage>45</fpage>
</bibl>

<bibl id="B34">
  <title><p>How the world's collective attention is being paid to a pandemic:
  COVID-19 related n-gram time series for 24 languages on Twitter</p></title>
  <aug>
    <au><snm>Alshaabi</snm><fnm>T.</fnm></au>
    <au><snm>Minot</snm><fnm>J. R.</fnm></au>
    <au><snm>Arnold</snm><fnm>M. V.</fnm></au>
    <au><snm>Adams</snm><fnm>J. L.</fnm></au>
    <au><snm>Dewhurst</snm><fnm>D. R.</fnm></au>
    <au><snm>Reagan</snm><fnm>A. J.</fnm></au>
    <au><snm>Muhamad</snm><fnm>R.</fnm></au>
    <au><snm>Danforth</snm><fnm>C. M.</fnm></au>
    <au><snm>Dodds</snm><fnm>P. S.</fnm></au>
  </aug>
  <source>\url{https://arxiv.org/pdf/2003.12614.pdf}. Accessed December 2,
  2020.</source>
</bibl>

<bibl id="B35">
  <title><p>Divergent modes of online collective attention to the COVID-19
  pandemic are associated with future caseload variance</p></title>
  <aug>
    <au><snm>Dewhurst</snm><fnm>DR</fnm></au>
    <au><snm>Alshaabi</snm><fnm>T</fnm></au>
    <au><snm>Arnold</snm><fnm>MV</fnm></au>
    <au><snm>Minot</snm><fnm>JR</fnm></au>
    <au><snm>Danforth</snm><fnm>CM</fnm></au>
    <au><snm>Dodds</snm><fnm>PS</fnm></au>
  </aug>
  <source>\url{https://arxiv.org/pdf/2004.03516.pdf}. Accessed December 2,
  2020.</source>
</bibl>

<bibl id="B36">
  <title><p>Weibo Reports First Quarter 2020 Unaudited Financial
  Results</p></title>
  <source>\url{http://ir.weibo.com/news-releases/news-release-details/weibo-reports-first-quarter-2020-unaudited-financial-results/}.
  Accessed December 2, 2020.</source>
</bibl>

<bibl id="B37">
  <title><p>An {Introduction} to {Sina} {Weibo} for {Journalists}</p></title>
  <aug>
    <au><snm>Wang</snm><fnm>Y</fnm></au>
  </aug>
  <source>\url{https://www.interhacktives.com/2018/02/22/how-to-use-sina-weibo-as-a-journalist/}.
  Accessed December 2, 2020.</source>
</bibl>

<bibl id="B38">
  <title><p>Common Questions on the Rules of Real-time Hot-Search-List,
  Hot-Message-List and Hot-Topic-List</p></title>
  <aug>
    <au><snm>Service</snm><fnm>WC</fnm></au>
  </aug>
  <source>\url{https://www.weibo.com/ttarticle/p/show?id=2309404007731978739654}.
  Accessed December 2, 2020.</source>
</bibl>

<bibl id="B39">
  <title><p>Weibo {Advertising}</p></title>
  <source>\url{https://www.marketingtochina.com/weibo-advertising/}. Accessed
  December 2, 2020.</source>
</bibl>

<bibl id="B40">
  <title><p>National Health Commission of People's Republic of
  China</p></title>
  <source>\url{http://www.nhc.gov.cn/xcs/xxgzbd/gzbd_index.shtml}. Accessed
  December 2, 2020.</source>
</bibl>

<bibl id="B41">
  <title><p>{China confirms 15152 new coronavirus cases, 254 additional
  deaths}</p></title>
  <source>\url{https://www.cnbc.com/2020/02/13/coronavirus-latest-updates-china-hubei.html}.
  Accessed December 2, 2020.</source>
</bibl>

<bibl id="B42">
  <title><p>China reports 99 new virus cases, majority imported</p></title>
  <aug>
    <au><snm>Sajid</snm><fnm>I</fnm></au>
  </aug>
  <source>\url{https://www.aa.com.tr/en/asia-pacific/china-reports-99-new-virus-cases-majority-imported/1801667}.
  Accessed December 2, 2020.</source>
</bibl>

<bibl id="B43">
  <title><p>Savitzky–Golay filter</p></title>
  <source>\url{https://en.wikipedia.org/wiki/Savitzky-Golay_filter}. Accessed
  December 2, 2020.</source>
</bibl>

</refgrp>
} 


\begin{thebibliography}{9}
	
\bibitem{TwitterTrend} 
Twitter Trend. 
{\color{blue}{https://trends24.in/about}}. Accessed August 8, 2020.

\bibitem{Zubiaga2011} 
Zubiaga, A., Spina, D., Fresno, V., Martínez, R.: Classifying trending topics: A typology of conversation triggers on twitter. In: Berendt, B., de Vries, A., Fan, W. (eds.) Proceedings of the 20th ACM International Conference on Information and Knowledge Management, Glasgow, UK (2011). ACM, New York

\bibitem{Lee_etal2011} 
Lee, K., Palsetia, D., Narayanan, R., Patwary, M.A., Agrawal, A., Choudhary, A.: Twitter trending topic classification. In: Spiliopoulou, M., et al. (eds.) Proceedings of the 11th IEEE International Conference on Data Mining Workshops, Vancouver, Canada, pp. 251–258 (2011). IEEE Computer Society, Los Alamitos

\bibitem{Tam2020} 
Tam, S., Hahm, D.: Exploring Coronavirus Twitter Trends. {\color{blue}{https://towardsdatascience.com/coronavirus-twitter-trends-d32fed5a027e}}. Accessed August 8, 2020. Towards Data Science, Inc. (2020)

\bibitem{Twitter-Covid} 
COVID-19 stream. {\color{blue}{https://developer.twitter.com/en/docs/labs/covid19-stream/overview}}. Accessed August 7, 2020. Twitter, Inc. (2020)

\end{thebibliography}
